\documentclass[journal]{IEEEtran}
\usepackage{amsmath}
\usepackage{amssymb}
\usepackage{amsfonts}
\usepackage{slashbox}
\usepackage{graphicx}
\usepackage{cite, amsmath, amsfonts, amssymb, psfrag, epsfig,tikz}
\usetikzlibrary{shapes,shadows,arrows}
\newtheorem{proposition}{{Proposition}}
\newtheorem{definition}{{Definition}}
\newtheorem{theorem}{{Theorem}}

\newtheorem{remark}{{Remark}}

\DeclareMathAlphabet{\mathpzc}{OT1}{pzc}{m}{it}
\ifCLASSINFOpdf
\else
\fi
\begin{document}
%
\title{Generalized Gaussian Multiterminal Source Coding: The Symmetric Case}
%
%
%

\author{Jun Chen, Li Xie, Yameng Chang, Jia Wang, Yizhong Wang
}

\maketitle

\begin{abstract}
Consider a generalized multiterminal source coding system, where $\ell\choose m$ encoders, each observing a distinct size-$m$ subset of
$\ell$ ($\ell\geq 2$) zero-mean unit-variance symmetrically correlated Gaussian sources with correlation coefficient $\rho$, compress their observations in such a way that a joint
decoder can reconstruct the sources within a prescribed mean squared error distortion based on the compressed data. The optimal rate-distortion performance of this system was previously known only for the two extreme cases $m=\ell$ (the centralized case) and $m=1$ (the distributed case), and except when $\rho=0$, the centralized system can achieve strictly lower compression rates than the distributed system under all non-trivial distortion constraints.
Somewhat surprisingly, it is established in the present paper that the optimal rate-distortion performance of the afore-described generalized multiterminal source coding system with $m\geq 2$ coincides with that of the centralized system for all distortions when $\rho\leq 0$ and for distortions below an explicit positive threshold (depending on $m$) when $\rho>0$. Moreover, when $\rho>0$, the minimum achievable rate of generalized multiterminal source coding subject to an arbitrary positive distortion constraint $d$ is shown to be within a finite gap (depending on $m$ and $d$) from its centralized counterpart in the large $\ell$ limit except for possibly  the critical distortion $d=1-\rho$.
\end{abstract}


%
\IEEEpeerreviewmaketitle

\section{Introduction}
%
%
%
%
\IEEEPARstart{M}{ultiterminal} source coding deals with the scenarios where (possibly) correlated data collected at different sites are compressed in a distributed manner and then forwarded to a fusion center for joint reconstruction. The fundamental problem here is to characterize the optimal tradeoff between the compression rates and the reconstruction distortions. The lossless version of this problem was largely solved by Slepian and Wolf in their landmark paper \cite{SW73}. Their result was later partially extended to the lossy case by Wyner and Ziv \cite{WZ76} and by Berger and Tung \cite{Berger78, Tung78}. Though a complete solution to the general lossy multiterminal source coding problem remains out of reach, significant progress has been made on some special cases of this problem, most notably the quadratic Gaussian case \cite{Oohama97, WTV08, WCW10, YZX13, WC13, WC14, Oohama14} and the logarithmic loss case \cite{CW14}.

\begin{figure}[tb]
\hspace{0.19in}
\includegraphics[width=7.6cm]{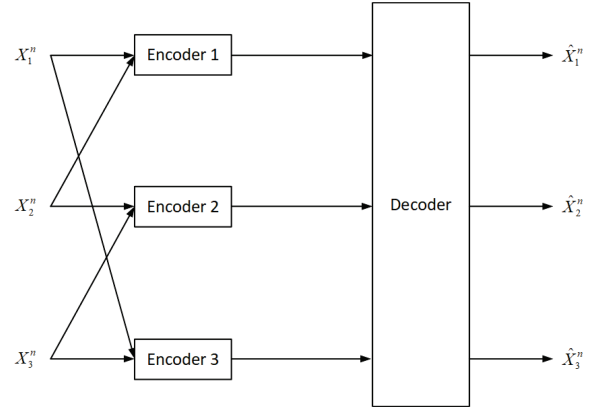}
\caption{A generalized multiterminal source coding system with $(\ell,m)=(3,2)$. \label{fig:plot1}}
\end{figure}

In many applications, the data collected at one site may be partially contained in those collected at another site.
For example, in a distributed video surveillance system, the scenes captured by different cameras can potentially overlap with each other.  To model such scenarios, a so-called generalized multiterminal source coding problem was introduced in \cite{CEK17}. Specifically, in generalized multiterminal source coding, several encoders, each observing a subset of $\ell$ jointly distributed sources, compress their observations in such a way that a joint
decoder can reconstruct the sources within a prescribed distortion level based on the compressed data. It is shown in \cite{CEK17} that, for Gaussian sources with mean squared error distortion constraints, a generalized multiterminal source coding system can achieve the same rate-distortion performance as that of the centralized point-to-point system in the high-resolution regime if the source-encoder bipartite graph and the probabilistic graphical model of the source distribution satisfy a certain condition.

In this work, we shall continue this line of research by considering a symmetric version of the generalized Gaussian multiterminal source coding problem. Here we have $\ell$ zero-mean unit-variance symmetrically correlated Gaussian sources with correlation coefficient $\rho$ and ${\ell\choose m}$ encoders, each of which has access to a distinct size-$m$ subset of these $\ell$ sources (see Fig. \ref{fig:plot1} for an illustration of the case $(\ell,m)=(3,2)$); moreover, we impose a normalized mean squared error trace distortion constraint on the joint source reconstruction (or equivalently, identical mean squared error distortion constraints on individual source reconstructions). It is worth mentioning that this seemingly simple symmetric setting is in fact non-trivial. Indeed, the associated rate-distortion function was previously known only for the two extreme cases $m=\ell$ (the centralized case) and $m=1$ (the distributed case). Furthermore, there are two major benefits to study this symmetric setting. First of all, it enables us to obtain results that are more explicit and conclusive than those for a more generic setting in \cite{CEK17}. More importantly, it is instructive to think of $m$ as a parameter that specifies the amount of cooperation among the encoders; as such, one can gain a precise understanding of the value of cooperation in terms of improving compression efficiency by investigating the gradual transition from a distributed system to a centralized system with $m$ varying from 1 to $\ell$.





The rest of this paper is organized as follows. We provide the problem definition and the statement of the main results in Section \ref{sec:main}. The proofs of the mains results can be found in Sections \ref{sec:proof1}, \ref{sec:proof2}, and \ref{sec:proof3}. We present some numerical results in Section \ref{sec:numerical}. Section \ref{sec:conclusion} contains the concluding remarks.

Notation: We use $\mathbb{E}[\cdot]$, $(\cdot)^T$, $\mathrm{tr}(\cdot)$, and $\det(\cdot)$  to denote the expectation operator, the transpose operator,  the trace operator, and the determinant operator, respectively. For any random (column) vector $Y$ and random object $\omega$, the  distortion covariance matrix incurred by the minimum mean squared error estimator of $Y$ from $\omega$ (i.e., $\mathbb{E}[(Y-\mathbb{E}[Y|\omega])((Y-\mathbb{E}[Y|\omega]))^T]$) is denoted by $\mathrm{cov}(Y|\omega)$.
We use $Y^n$ as an abbreviation of $(Y(1),\cdots,Y(n))$. The cardinality of a set $\mathcal{S}$ is denoted by $|\mathcal{S}|$. An $\ell\times\ell$ diagonal matrix with the $i$-th diagonal entry being $a_i$, $i=1,\cdots,\ell$, is written as $\mathrm{diag}(a_1,\cdots,a_{\ell})$. Throughout this paper, the base of the logarithm function is $e$.

\section{Problem Definition and Main Results}\label{sec:main}

Let $X\triangleq(X_1,\cdots,X_{\ell})^T$ be an $\ell$-dimensional ($\ell\geq 2$) zero-mean Gaussian random column vector with covariance matrix
\begin{align*}
\Sigma^{(\ell)}=\left(
  \begin{array}{cccc}
    1 & \rho & \cdots &  \rho \\
    \rho & \ddots & \ddots &   \vdots \\
    \vdots & \ddots & \ddots & \rho \\
     \rho & \cdots & \rho & 1 \\
  \end{array}
\right).
\end{align*}
We assume $\rho\in(-\frac{1}{\ell-1},1)$ to ensure that $\Sigma^{(\ell)}$ is positive definite. Let $X(t)\triangleq(X_1(t),\cdots,X_{\ell}(t))^T$, $t=1,2,\cdots$, be i.i.d. copies of $X$.

\begin{definition}
A rate $r$ is said to be achievable by an $(\ell,m)$ generalized multiterminal source coding system under normalized mean squared error trace distortion constraint $d$ if, for any $\epsilon>0$, there exist encoding functions $\phi^{(n)}_{\mathcal{S}}:\mathbb{R}^{m\times n}\rightarrow\mathcal{C}^{(n)}_{\mathcal{S}}$, $\mathcal{S}\in\mathcal{I}^{(\ell,m)}\triangleq\{\mathcal{S}\subseteq\{1,\cdots,\ell\}:|\mathcal{S}|=m\}$, and a decoding function $\psi^{(n)}:\prod_{\mathcal{S}\in\mathcal{I}^{(\ell,m)}}\mathcal{C}^{(n)}_{\mathcal{S}}\rightarrow\mathbb{R}^{\ell\times n}$ such that
\begin{align}
&\frac{1}{n}\sum\limits_{\mathcal{S}\in\mathcal{I}^{(\ell,m)}}\log|\mathcal{C}^{(n)}_{\mathcal{S}}|\leq r+\epsilon,\nonumber\\
&\frac{1}{\ell n}\sum\limits_{t=1}^n\mathrm{tr}(\mathbb{E}[(X(t)-\hat{X}(t))(X(t)-\hat{X}(t))^T])\leq d+\epsilon,\label{eq:tracedistortion}
\end{align}
where
\begin{align*}
\hat{X}^n\triangleq\psi^{(n)}(\phi^{(n)}_{\mathcal{S}}(X^n_i, i\in\mathcal{S}), \mathcal{S}\in\mathcal{I}^{(\ell,m)}).
\end{align*}
The minimum of such $r$ is denoted by $r^{(\ell,m)}(d)$, which will referred to as the rate-distortion function of $(\ell,m)$ generalized multiterminal source coding.
\end{definition}
\begin{remark}
Due to the symmetry of the source distribution, $r^{(\ell,m)}(d)$ remains the same if we replace the normalized mean squared error trace distortion constraint on the joint source reconstruction in (\ref{eq:tracedistortion}) with identical mean squared error distortion constraints on individual source reconstructions given below
\begin{align*}
\frac{1}{n}\sum\limits_{t=1}^n\mathbb{E}[(X_i(t)-\hat{X}_i(t))^2]\leq d+\epsilon,\quad i=1,\cdots,\ell,
\end{align*}
where $\hat{X}_i(t)$ is the $i$-th entry of $\hat{X}(t)$, $i=1,\cdots,\ell$, $t=1,\cdots,n$.
\end{remark}

\begin{remark}
It is clear that, for $m=1,\cdots,\ell$,
\begin{align*}
r^{(\ell,m)}(d)=0,\quad d\geq 1.
\end{align*}
Henceforth we shall assume $d\in(0,1)$.
\end{remark}

\begin{remark}\label{rmk:remark3}
Note that  an encoder that observes $X^n_i$, $i\in\mathcal{S}$, is at least as powerful as one that observes $X^n_i$, $i\in\mathcal{S}'$, for some $\mathcal{S}'\subseteq\mathcal{S}$, in the sense that the former can perform any function that the latter can do. Given $1\leq m'<m\leq\ell$, we can find, for any $(\ell, m')$ generalized multiterminal source coding system, an $(\ell,m)$ generalized multiterminal source coding system such that each encoder in the $(\ell, m')$ system is dominated (in terms of functionality) by an encoder in the $(\ell,m)$ system. Therefore, we must have $r^{(\ell,m)}(d)\leq r^{(\ell,m')}(d)$ for $m> m'$.
\end{remark}

A complete characterization of $r^{(\ell,m)}(d)$ was previously known only for $m=\ell$ and $m=1$. It is instructive to review the relevant results for these two extreme cases since they provide the necessary background and useful motivations for the introduction of our new results.

First recall the following results, which can be specialized from the general theory of circulant matrices \cite{Gray06}.
For any $\ell\times\ell$ real matrix $\Pi$ of the form
\begin{align}
\left(
  \begin{array}{cccc}
    a & b & \cdots &  b \\
    b & \ddots & \ddots &   \vdots \\
    \vdots & \ddots & \ddots & b \\
     b & \cdots & b & a \\
  \end{array}
\right),\label{eq:form}
\end{align}
its eigenvalues are given by
\begin{align}
&\lambda_i\triangleq a-b,\quad i=1,\cdots,\ell-1,\label{eq:cireigen1}\\
&\lambda_{\ell}\triangleq a+(\ell-1)b,\label{eq:cireigen2}
\end{align}
and we have
\begin{align*}
\det(\Pi)=\prod\limits_{i=1}^{\ell}\lambda_i=(a-b)^{\ell-1}(a+(\ell-1)b).
\end{align*}
The normalized eigenvectors corresponding to $\lambda_1,\cdots,\lambda_{\ell}$ can be constructed in such a way that they are orthogonal to each other and do not depend on $a$ and $b$. Typically these eigenvectors are chosen to be the Fourier basis, but it is also possible to construct the real ones. The exact form of these eigenvectors are inessential for our purpose.
It will be seen that the source covariance matrix and the distortion covariance matrices encountered in this work are all of the form (\ref{eq:form}); as a consequence, they can all be diagonalized by the same unitary matrix. Note that, in an $(\ell,m)$ generalized multiterminal source coding system with $m<\ell$, each encoder can only observe a subset of the sources; therefore, in principle it cannot decorrelate the sources simultaneously through a unitary transformation and perform compression in the transform domain (i.e., the eigenspace). Nevertheless, due to the special form of the resulting distortion covariance matrix, one may still interpret the effect of such a system and make sensible comparisons with that of the centralized system (i.e., $m=\ell$) in the transform domain.

For reasons that will become clear soon, we define
\begin{align*}
&d^-_c\triangleq 1+(\ell-1)\rho,\\
&d^+_c\triangleq 1-\rho,
\end{align*}
and refer to them as critical distortions. It will be seen that these two critical distortions are of special importance.






Now consider the case $m=\ell$. One can determine $r^{(\ell,\ell)}(d)$ by solving the following convex optimization problem
\begin{align}
r^{(\ell,\ell)}(d)&=\min\limits_{D}\frac{1}{2}\log\frac{\det(\Sigma^{(\ell)})}{\det(D)}\label{eq:convex}\\
\mbox{subject to }&\quad 0\prec D\preceq\Sigma^{(\ell)},\nonumber\\
&\quad\frac{1}{\ell}\mathrm{tr}(D)\leq d,\nonumber
\end{align}
where $A\prec(\preceq) B$ means $B-A$ is positive (semi)definite. The optimal solution to this minimization problem is unique and is given by
\begin{align*}
D=D^{(\ell,\ell)}\triangleq\left(
  \begin{array}{cccc}
    d & \theta^{(\ell,\ell)} & \cdots & \theta^{(\ell,\ell)} \\
    \theta^{(\ell,\ell)} & \ddots & \ddots &   \vdots \\
    \vdots & \ddots & \ddots & \theta^{(\ell,\ell)} \\
     \theta^{(\ell,\ell)} & \cdots & \theta^{(\ell,\ell)} & d \\
  \end{array}
\right),
\end{align*}
where, for $\rho\in(\frac{1}{\ell-1},0]$,
\begin{align}
\theta^{(\ell,\ell)}\triangleq\left\{
                     \begin{array}{ll}
                       0, & d\in(0, d^-_c),\\
                       \frac{1-d}{\ell-1}+\rho, & d\in[d^-_c,1),
                     \end{array}
                   \right.\label{eq:thetaellneg}
\end{align}
and, for $\rho\in(0,1)$,
\begin{align}
\theta^{(\ell,\ell)}\triangleq\left\{
                     \begin{array}{ll}
                       0, & d\in(0, d^+_c), \\
                        d-1+\rho, & d\in[d^+_c,1).
                     \end{array}
                   \right.\nonumber
\end{align}

An alternative approach is to solve the problem in the eigenspace.
Let $\lambda^{(\ell)}_1,\cdots,\lambda^{(\ell)}_{\ell}$ be the eigenvalues of $\Sigma^{(\ell)}$. It follows from (\ref{eq:cireigen1}) and (\ref{eq:cireigen2}) that
\begin{align}
&\lambda^{(\ell)}_i=1-\rho,\quad i=1,\cdots,\ell-1,\label{eq:eigen1}\\
&\lambda^{(\ell)}_{\ell}=1+(\ell-1)\rho.\label{eq:eigen2}
\end{align}
Note that the smallest eigenvalue coincides with $d^-_c$ for $\rho\in(-\frac{1}{\ell-1},0]$ and coincides with $d^+_c$ for $\rho\in(0,1)$.
One can determine $r^{(\ell,\ell)}(d)$ by solving the following distortion allocation problem
\begin{align}
r^{(\ell,\ell)}(d)&=\min\limits_{d_1,\cdots,d_{\ell}}\sum\limits_{i=1}^{\ell}\frac{1}{2}\log\frac{\lambda^{(\ell)}_i}{d_i}\label{eq:eigendistortion}\\
\mbox{subject to }&\quad 0< d_i\leq\lambda^{(\ell)}_i,\quad i=1,\cdots,\ell,\nonumber\\
&\quad\frac{1}{\ell}\sum\limits_{i=1}^nd_i\leq d.\nonumber
\end{align}
Its optimal solution is unique and is given by the well-known reverse water-filling formula \cite[Thm. 13.3.3]{CT91}
\begin{align}
d_i=d^{(\ell,\ell)}_{i}\triangleq\left\{
      \begin{array}{ll}
        \tilde{d}, & \tilde{d}<\lambda^{(\ell)}_i, \\
        \lambda_i, &  \tilde{d}\geq\lambda^{(\ell)}_i,
      \end{array}
    \right.\quad i=1,\cdots,\ell,\label{eq:distortionallocation}
\end{align}
with $\tilde{d}$ chosen such that $\frac{1}{\ell}\sum_{i=1}^{\ell}d^{(\ell,\ell)}_{i}=d$.
Substituting (\ref{eq:eigen1}) and (\ref{eq:eigen2}) into (\ref{eq:distortionallocation}) gives, for $\rho\in(-\frac{1}{\ell-1},0]$,
\begin{align*}
&d^{(\ell,\ell)}_i=\left\{
                    \begin{array}{ll}
                      d, & d\in(0, d^-_c), \\
                      \frac{\ell d-1}{\ell-1}-\rho, & d\in[d^-_c,1),
                    \end{array}
                  \right.\quad i=1,\cdots,\ell-1,\\
&d^{(\ell,\ell)}_{\ell}=\left\{
                         \begin{array}{ll}
                           d, & d\in(0, d^-_c), \\
                           1+(\ell-1)\rho, & d\in[d^-_c,1),
                         \end{array}
                       \right.
\end{align*}
and, for $\rho\in(0,1)$,
\begin{align*}
&d^{(\ell,\ell)}_i=\left\{
                    \begin{array}{ll}
                      d, & d\in(0, d^+_c), \\
                      1-\rho, & d\in[d^+_c,1),
                    \end{array}
                  \right.\quad i=1,\cdots,\ell-1,\\
&d^{(\ell,\ell)}_{\ell}=\left\{
                         \begin{array}{ll}
                           d, & d\in(0, d^+_c), \\
                           \ell d-(\ell-1)(1-\rho), & d\in[d^+_c,1).
                         \end{array}
                       \right.
\end{align*}
Note that $d^{(\ell,\ell)}_1,\cdots,d^{(\ell,\ell)}_{\ell}$ are exactly the eigenvalues of $D^{(\ell,\ell)}$.

It can be readily seen that both approaches lead to the following result.

\begin{proposition}\label{prop:ell}
For $\rho\in(-\frac{1}{\ell-1},0]$,
\begin{align*}
r^{(\ell,\ell)}(d)=\left\{
         \begin{array}{ll}
           \frac{1}{2}\log\frac{(1-\rho)^{\ell-1}(1+(\ell-1)\rho)}{d^{\ell}}, & d\in(0, d^-_c), \\
           \frac{\ell-1}{2}\log\frac{(\ell-1)(1-\rho)}{\ell d-1-(\ell-1)\rho}, & d\in[d^-_c,1).
         \end{array}
       \right.
\end{align*}
For $\rho\in(0,1)$,
\begin{align*}
r^{(\ell,\ell)}(d)=\left\{
         \begin{array}{ll}
           \frac{1}{2}\log\frac{(1-\rho)^{\ell-1}(1+(\ell-1)\rho)}{d^{\ell}}, & d\in(0, d^+_c), \\
           \frac{1}{2}\log\frac{1+(\ell-1)\rho}{\ell d-(\ell-1)(1-\rho)}, & d\in[d^+_c,1).
         \end{array}
       \right.
\end{align*}
\end{proposition}

It is easy to show from (\ref{eq:convex}) using
Hadamard's inequality and the arithmetic-geometric means inequality (or from (\ref{eq:eigendistortion}) using the arithmetic-geometric means inequality) that
\begin{align*}
r^{(\ell,\ell)}\geq\underline{r}^{(\ell)}(d)\triangleq\frac{1}{2}\log\frac{(1-\rho)^{\ell-1}(1+(\ell-1)\rho)}{d^L}.
\end{align*}
We shall refer to $\underline{r}^{(\ell)}(d)$ as the Shannon lower bound. Proposition \ref{prop:ell} indicates that $r^{(\ell,\ell)}(d)$ coincides with $\underline{r}^{(\ell)}(d)$ when $ d\in(0, d^-_c]$ for $\rho\in(-\frac{1}{\ell-1},0]$, and when $d\in(0, d^+_c]$ for $\rho\in(0,1)$.



Next consider the other extreme case $m=1$. The following result was first proved in \cite{WTV08} for $\rho\in[0,1)$ and then in \cite{WCW10} for $\rho\in(-\frac{1}{\ell-1},1)$. 

\begin{proposition}\label{prop:one}
For $\rho\in(-\frac{1}{\ell-1},1)$,
\begin{align*}
&r^{(\ell,1)}(d)\\
&=\frac{1}{2}\log\frac{(1-\rho)^{\ell-1}(1+(\ell-1)\rho)}{(d-\theta^{(\ell,1)})^{\ell-1}(d+(\ell-1)\theta^{(\ell,1)})},\quad d\in(0,1),
\end{align*}
where
\begin{align*}
\theta^{(\ell,1)}\triangleq\frac{\rho d\gamma^{(\ell,1)}}{\gamma^{(\ell,1)}+(1-\rho)(1+(\ell-1)\rho)}
\end{align*}
with
\begin{align*}
&\gamma^{(\ell,1)}\triangleq\frac{-\xi+\sqrt{\xi^2+4(1-\rho)(1+(\ell-1)\rho)d(1-d)}}{2(1-d)},\\
&\xi\triangleq(1+(\ell-1)\rho)(1-\rho-d)-(1-\rho)d.
\end{align*}
\end{proposition}

To understand its connection with $r^{(\ell,\ell)}(d)$, it is instructive to write $r^{(\ell,1)}(d)$ as
\begin{align*}
r^{(\ell,1)}(d)=\frac{1}{2}\log\frac{\det(\Sigma^{(\ell)})}{\det(D^{(\ell,1)})},
\end{align*}
where
\begin{align*}
D^{(\ell,1)}\triangleq\left(
  \begin{array}{cccc}
    d & \theta^{(\ell,1)} & \cdots & \theta^{(\ell,1)} \\
    \theta^{(\ell,1)} & \ddots & \ddots &   \vdots \\
    \vdots & \ddots & \ddots & \theta^{(\ell,1)} \\
     \theta^{(\ell,1)} & \cdots & \theta^{(\ell,1)} & d \\
  \end{array}
\right).
\end{align*}
We can also express $r^{(\ell,1)}(d)$ alternatively as
\begin{align*}
r^{(\ell,1)}(d)=\sum\limits_{i=1}^{\ell}\frac{1}{2}\log\frac{\lambda^{(\ell)}_i}{d^{(\ell,1)}_i},
\end{align*}
where
\begin{align*}
&d^{(\ell,1)}_i\triangleq d-\theta^{(\ell,1)},\quad i=1,\cdots,\ell-1,\\
&d^{(\ell,1)}_{\ell}\triangleq d+(\ell-1)\theta^{(\ell,1)},
\end{align*}
are the eigenvalues of $D^{(\ell,1)}$. It can be verified that $D^{(\ell,1)}\neq D^{(\ell,\ell)}$ and $(d^{(\ell,1)}_1,\cdots,d^{(\ell,1)}_{\ell})\neq (d^{(\ell,\ell)}_1,\cdots,d^{(\ell,\ell)}_{\ell})$ unless $\rho=0$. Therefore, we must have, for $\rho\in(-\frac{1}{\ell-1},0)\cup(0,1)$,
\begin{align*}
r^{(\ell,1)}(d)>r^{(\ell,\ell)}(d),\quad d\in(0,1).
\end{align*}

One might be inclined to expect that $r^{(\ell,m)}(d)$ is strictly greater than $r^{(\ell,\ell)}(d)$ for any $m<\ell$ unless the sources are independent or the distortion constraint is trivial. Somewhat surprisingly, it was shown in \cite{CEK17} that, in the high-resolution regime (i.e.,  when $d$ is sufficiently close to zero), $r^{(\ell,m)}(d)$ coincides with $r^{(\ell,\ell)}(d)$  when $m\geq 2$. However, the high-resolution condition in  \cite{CEK17} is not explicit. Our first main result shows that this high-resolution condition is in fact redundant  when the correlation coefficient $\rho$ is non-positive.

\begin{theorem}\label{thm:theorem1}
For $\rho\in(-\frac{1}{\ell-1},0]$ and $m=2,\cdots,\ell$,
\begin{align*}
r^{(\ell,m)}(d)=r^{(\ell,\ell)}(d),\quad d\in(0,1).
\end{align*}
\end{theorem}
\begin{IEEEproof}
See Section \ref{sec:proof1}
\end{IEEEproof}

For positive $\rho$, we have the following result, which provides an explicit high-resolution condition under which $r^{(\ell,m)}(d)$ (with $m\geq 2$) matches $r^{(\ell,\ell)}(d)$.
\begin{theorem}\label{thm:theorem2}
For $\rho\in(0,1)$ and $m=1,\cdots,\ell$,
\begin{align*}
r^{(\ell,m)}(d)=r^{(\ell,\ell)}(d),\quad d\in(0,d^{(\ell,m)}_c],
\end{align*}
where
\begin{align*}
d^{(\ell,m)}_c\triangleq 1-\frac{(\ell-1)\rho(1+(m-1)\rho)}{(\ell-1)m\rho+(m-1)(1-\rho)}.
\end{align*}
\end{theorem}
\begin{IEEEproof}
See Section \ref{sec:proof2}.
\end{IEEEproof}
\begin{remark}
We have $d^{(\ell,\ell)}_c=d^+_c$ and $d^{(\ell,1)}_c=0$. The statement of Theorem \ref{thm:theorem2}  is trivial when $m=\ell$ and is void when $m=1$.
\end{remark}
\begin{remark}
$d^{(\ell,m)}_c$ is a monotonically increasing function of $m$ for fixed $\ell$ and is a monotonically decreasing function of $\ell$ for fixed $m$. Moreover, we have
\begin{align*}
&\lim\limits_{\ell\rightarrow\infty}d^{(\ell,m)}_c=d^{(m)}_c\triangleq\frac{(m-1)(1-\rho)}{m},\\
&\lim\limits_{m\rightarrow\infty}d^{(m)}_c=d^+_c,
\end{align*}
which implies that, for $\rho\in(0,1)$, $r^{(\ell,m)}(d)$ essentially matches  $r^{(\ell,\ell)}(d)$ (and the Shannon lower bound $\underline{r}^{(\ell)}(d)$ as well) all the way up to the critical distortion $d^+_c$ when $\ell$ and $m$ are sufficiently large (even if the ratio $\frac{m}{\ell}$ is close to zero).
\end{remark}

It remains to understand the behavior of $r^{(\ell,m)}(d)$  when $d>d^{(\ell,m)}_c$ for $\rho\in(0,1)$ and $m\geq 2$. To simplify the analysis, we shall consider the asymptotic regime where $\ell$ goes to infinity with $m$ fixed. Define
\begin{align*}
r^{(\ell,m)}_1(d)&\triangleq\frac{\ell}{2}\log\frac{1-\rho}{d}+\frac{1}{2}\log\ell+\frac{1}{2}\log\frac{\rho}{1-\rho}+O(\frac{1}{\ell}),\\
r^{(\ell,m)}_2(d)&\triangleq\frac{\ell}{2}\log\frac{1-\rho}{d}+\frac{1}{2}\log\ell\\
&\quad+\frac{d-(m-1)(1-\rho-d)}{2m(1-\rho-d)}\\
&\quad+\frac{1}{2}\log\frac{m\rho(1-\rho-d)}{(1-\rho)^2}+O(\frac{1}{\ell}),\\
r^{(\ell,m)}_3(d)&\triangleq\frac{\sqrt{\ell}}{2\sqrt{m}}+\frac{1}{4}\log\ell+\frac{1}{2}\log\frac{\sqrt{m}\rho}{1-\rho}\\
&\quad-\frac{1+(m-1)\rho}{4m\rho}+O(\frac{1}{\sqrt{\ell}}),\\
r^{(\ell,m)}_4(d)&\triangleq\frac{1}{2}\log\frac{\rho}{d-1+\rho}+\frac{(1-\rho)(1-d)}{2m\rho(d-1+\rho)}+O(\frac{1}{\ell}),
\end{align*}
where $g(\ell)=O(f(\ell))$ means the absolute value of $\frac{g(\ell)}{f(\ell)}$ is bounded for all sufficiently large $\ell$.

\begin{theorem}\label{thm:theorem3}
For $\rho\in(0,1)$ and $m\geq 1$,
\begin{align*}
&r^{(\ell,m)}(d)\leq\left\{
                       \begin{array}{ll}
                         r^{(\ell,m)}_1(d), & d\in(0,d^{(m)}_c], \\
                        r^{(\ell,m)}_2(d), & d\in(d^{(m)}_c,d^+_c), \\
                       r^{(\ell,m)}_3(d) , & d=d^+_c,\\
                        r^{(\ell,m)}_4(d),  & d\in(d^+_c,1).
                       \end{array}
                     \right.
\end{align*}
Moreover, this upper bound is tight when $m=1$ or $d\in(0,d^{(m)}_c]$.
\end{theorem}
\begin{IEEEproof}
See Section \ref{sec:proof3}.
\end{IEEEproof}

\begin{remark}
It follows from Proposition \ref{prop:ell} that, for $\rho\in(0,1)$,
\begin{align}
&r^{(\ell,\ell)}(d)\nonumber\\
&=\left\{
                        \begin{array}{ll}
                          \frac{\ell}{2}\log\frac{1-\rho}{d}+\frac{1}{2}\log\ell+\frac{1}{2}\log\frac{\rho}{1-\rho}+O(\frac{1}{\ell}), & d\in(0,d^+_c), \\
                          \frac{1}{2}\log\ell+\frac{1}{2}\log\frac{\rho}{1-\rho}+O(\frac{1}{\ell}), & d=d^+_c, \\
                          \frac{1}{2}\log\frac{\rho}{d-1+\rho}+O(\frac{1}{\ell}), & d\in(d^+_c,1).
                        \end{array}
                      \right.\label{eq:expansionell}
\end{align}
Combining Theorem \ref{thm:theorem3} and (\ref{eq:expansionell}) shows that, for $\rho\in(0,1)$ and $m\geq 1$,
\begin{align*}
\limsup\limits_{\ell\rightarrow\infty}r^{(\ell,m)}(d)-r^{(\ell,\ell)}(d)\leq\delta^{(m)}(d),\quad d\in(0,1),
\end{align*}
where
\begin{align*}
&\delta^{(m)}(d)\\
&\triangleq\left\{
                       \begin{array}{ll}
                         0, & d\in(0,d^{(m)}_c],\\
                         \frac{1-\rho-m(1-\rho-d)}{2m(1-\rho-d)}+\frac{1}{2}\log\frac{m(1-\rho-d)}{1-\rho}, & d\in(d^{(m)}_c,d^+_c),\\
                         \infty, & d=d^+_c,\\
                         \frac{(1-\rho)(1-d)}{2m\rho(d-1+\rho)}, & d\in(d^+_c,1).
                       \end{array}
                     \right.
\end{align*}
Note that, as a function of $d$ (with $m$ fixed), $\delta^{(m)}(d)$ is monotonically increasing  for $d\in(0,d^+_c)$ and  monotonically decreasing for $d\in(d^+_c,1)$; moreover, it approaches infinity as $d\rightarrow d^+_c$. For fixed $d$, $\delta^{(m)}(d)$ is a monotonically decreasing function of $m$ and converges to zero (though not uniformly over $d$) as $m\rightarrow\infty$ except at $d=d^+_c$. Therefore, for $\rho\in(0,1)$, $r^{(\ell,m)}(d)$ is within a finite gap (depending on $d$) from $r^{(\ell,\ell)}(d)$ even in the limit of large $\ell$ when $d\neq d^+_c$; moreover, this gap diminishes as $m$ increases. For $\rho\in(0,1)$, the gap between $r^{(\ell,m)}(d^+_c)$ and $r^{(\ell,\ell)}(d^+_c)$ can potentially approaches infinity as $\ell\rightarrow\infty$, and is indeed so when $m=1$.
\end{remark}




\begin{remark}
In view of Theorem \ref{thm:theorem3}, (\ref{eq:expansionell}), and Remark \ref{rmk:remark3}, we have, for $\rho\in(0,1)$ and $m\geq 1$,
\begin{align*}
\lim\limits_{\ell\rightarrow\infty}\frac{1}{\ell}r^{(\ell,m)}(d)=\left\{
                                   \begin{array}{ll}
                                     \frac{1}{2}\log\frac{1-\rho}{d}, & d\in(0,d^+_c), \\
                                     0, & d\in[d^+_c,1),
                                   \end{array}
                                 \right.
\end{align*}
which implies that  the average minimum achievable rate per encoder of an $(\ell, m)$ generalized multiterminal source coding system is essentially independent of $m$ when $\ell$ is sufficiently large.
\end{remark}

\begin{remark}
It is interesting to see that, for $\rho\in(0,1)$ and $m\geq 1$, $r^{(\ell,m)}(d)$ remains bounded (though not uniformly over $d$) even in the limit of large $\ell$ when $d\in(d^+_c,1)$.
\end{remark}




\section{Proof of Theorem \ref{thm:theorem1}}\label{sec:proof1}

In view of Proposition \ref{prop:ell}, Proposition \ref{prop:one}, and Remark \ref{rmk:remark3}, for $\rho=0$ and $m=1,\cdots,\ell$, 
\begin{align*}
r^{(\ell,m)}(d)=\frac{\ell}{2}\log\frac{1}{d},\quad d\in(0,1).
\end{align*}
Therefore, we shall only consider the case $\rho\in(-\frac{1}{\ell-1},0)$.
It suffices to show that
\begin{align}
r^{(\ell,m)}(d)\leq r^{(\ell,\ell)}(d),\quad d\in(0,1),\label{eq:onedirectionnegative}
\end{align}
since the other direction is trivially true (see Remark \ref{rmk:remark3}).
To this end, we need the following result, which can be obtained by specializing  the well-known Berger-Tung upper bound \cite{Berger78, Tung78, XCWB07} to our current setting.
\begin{proposition}\label{prop:BergerTung}
For any Gaussian random variables/vectors $V_{\mathcal{S}}$, $\mathcal{S}\in\mathcal{I}^{(\ell,m)}$, jointly distributed with $X$ such that $V_{\mathcal{S}}\leftrightarrow(X_i,i\in\mathcal{S})\leftrightarrow(X_i', i'\in\{1,\cdots,\ell\}\backslash\mathcal{S}, V_{\mathcal{S}'},\mathcal{S}'\in\mathcal{I}^{(\ell,m)}\backslash\mathcal{S})$ form a Markov chain for any $\mathcal{S}\in\mathcal{I}^{(\ell,m)}$, we have
\begin{align*}
&r^{(\ell,m)}(\frac{1}{\ell}\mathrm{tr}(\mathrm{cov}(X|V_{\mathcal{S}},\mathcal{S}\in\mathcal{I}^{(\ell,m)})))\\
&\leq \frac{1}{2}\log\frac{\det(\Sigma^{(\ell)})}{\det(\mathrm{cov}(X|V_{\mathcal{S}},\mathcal{S}\in\mathcal{I}^{(\ell,m)}))}.
\end{align*}
\end{proposition}

Equipped with Proposition \ref{prop:BergerTung}, we are in a position to prove Theorem \ref{thm:theorem1}. Let $M$ be an $m\times m$ matrix given by
\begin{align*}
M\triangleq\left(
                                                \begin{array}{cccc}
                                                  m-1 & -1 & \cdots & -1 \\
                                                  -1 & \ddots & \ddots & \vdots \\
                                                  \vdots & \ddots & \ddots & -1 \\
                                                  -1 & \cdots & -1 & m-1 \\
                                                \end{array}
                                              \right).
\end{align*}
For any $\gamma>0$ and $\mathcal{S}\triangleq\{i_1,\cdots,i_m\}\in\mathcal{I}^{(\ell,m)}$ with $i_1<\cdots<i_m$, define
\begin{align*}
\left(
                           \begin{array}{c}
                             U^-_{\mathcal{S},1}(\gamma) \\
                             \vdots \\
                             \vdots \\
                             U^-_{\mathcal{S},m}(\gamma) \\
                           \end{array}
                         \right)
\triangleq  M\left(
                                                \begin{array}{c}
                                                  X_{i_1} \\
                                                  \vdots \\
                                                  \vdots \\
                                                  X_{i_m} \\
                                                \end{array}
                                              \right)+\sqrt{\gamma}\left(
                                                        \begin{array}{c}
                                                          N^-_{\mathcal{S},1} \\
                                                          \vdots \\
                                                          \vdots \\
                                                          N^-_{\mathcal{S},m} \\
                                                        \end{array}
                                                      \right),
\end{align*}
where $(N^-_{\mathcal{S},1},\cdots,N^-_{\mathcal{S},m})^T$ is a Gaussian random vector with mean zero and covariance matrix $M$. Moreover, we assume that $X$, $(N^-_{\mathcal{S},1},\cdots,N^-_{\mathcal{S},m})^T$, $\mathcal{S}\in\mathcal{I}^{(\ell,m)}$, are mutually independent.

\begin{proposition}\label{prop:negativecoefficient}
We have
\begin{align*}
&\mathrm{cov}(X|U^-_{\mathcal{S},1}(\gamma),\cdots,U^-_{\mathcal{S},m}(\gamma),\mathcal{S}\in\mathcal{I}^{(\ell,m)})\\
&=\left(
    \begin{array}{cccc}
      d^-(\gamma) & \theta^-(\gamma) & \cdots & \theta^-(\gamma) \\
      \theta^-(\gamma) & \ddots & \ddots & \vdots \\
      \vdots & \ddots & \ddots & \theta^-(\gamma) \\
      \theta^-(\gamma) & \cdots & \theta^-(\gamma) & d^-(\gamma) \\
    \end{array}
  \right),
\end{align*}
where
\begin{align*}
&d^-(\gamma)\triangleq 1-\frac{{\ell-2\choose m-2}(\ell-1)(1-\rho)^2}{\gamma+{\ell-2\choose m-2}\ell(1-\rho)},\\
&\theta^-(\gamma)\triangleq\rho+\frac{{\ell-2\choose m-2}(1-\rho)^2}{\gamma+{\ell-2\choose m-2}\ell(1-\rho)}.
\end{align*}
\end{proposition}
\begin{IEEEproof}
See Appendix \ref{app:negativecoefficient}.
\end{IEEEproof}

Setting $d^-(\gamma)=d$ gives
\begin{align*}
\gamma=\gamma^{(\ell,m)}\triangleq\frac{{\ell-2\choose m-2}(1-\rho)((\ell-1)(1-\rho)-\ell(1-d))}{1-d}.
\end{align*}
Note that there is a one-to-one correspondence between $d\in(\frac{d^-_c}{\ell},1)$ and $\gamma^{(\ell,m)}\in(0,\infty)$. Moreover,
\begin{align*}
\theta^-(\gamma^{(\ell,m)})=\frac{1-d}{\ell-1}+\rho,
\end{align*}
which coincides with $\theta^{(\ell,\ell)}$ in (\ref{eq:thetaellneg}) for $d\in[d^-_c,1)$; in particular, $\theta^-(\gamma^{(\ell,m)}_c)=0$, where
\begin{align*}
\gamma^{(\ell,m)}_c\triangleq-\frac{{\ell-2\choose m-2}(1-\rho)(1+(\ell-1)\rho)}{\rho}
\end{align*}
is the value of $\gamma^{(\ell,m)}$ at $d=d^-_c$.
Invoking Proposition \ref{prop:BergerTung} with $V_{\mathcal{S}}\triangleq(U^-_{\mathcal{S},1}(\gamma^{(\ell,m)}),\cdots,U^-_{\mathcal{S},m}(\gamma^{(\ell,m)}))^T$, $\mathcal{S}\in\mathcal{I}^{(\ell,m)}$, (which satisfy the Markov chain condition in Proposition \ref{prop:BergerTung}) proves (\ref{eq:onedirectionnegative}) for $d\in[d^-_c,1)$.

Now consider the case $d\in(0,d^-_c)$. Let
\begin{align*}
W^-_i(d)\triangleq X_i+\sqrt{\frac{d^-_cd}{d^-_c-d}}Z^-_i,\quad i=1,\cdots,\ell,
\end{align*}
where $Z^-_1,\cdots,Z^-_{\ell}$ are mutually independent zero-mean unit variance Gaussian random variables, and are independent of $X$, $(N^-_{\mathcal{S},1},\cdots,N^-_{\mathcal{S},m})^T$, $\mathcal{S}\in\mathcal{I}^{(\ell,m)}$. Construct $\Omega_{\mathcal{S}}$, $\mathcal{S}\in\mathcal{I}^{(\ell,m)}$, such that 1) $\Omega_{\mathcal{S}}\subseteq\mathcal{S}$, $\mathcal{S}\in\mathcal{I}^{(\ell,m)}$, 2) $\Omega_{\mathcal{S}}\cap\Omega_{\mathcal{S}'}=\emptyset$, $\mathcal{S}\neq\mathcal{S}'$, 3) $\cup_{\mathcal{S}\in\mathcal{I}^{(\ell,m)}}\Omega_{\mathcal{S}}=\{1,\cdots,\ell\}$. Such a construction always exists. For example, we can let
\begin{align*}
\Omega_{\mathcal{S}}\triangleq\left\{
                       \begin{array}{ll}
                         \mathcal{S}, & \mathcal{S}=\{1,\cdots,m\}, \\
                         \{i\}, & \mathcal{S}=\{i-m+1,\cdots,i\}, i=m+1,\cdots,\ell,  \\
                         \emptyset, & \mbox{otherwise}.
                       \end{array}
                     \right.
\end{align*}
Define $V_{\mathcal{S}}\triangleq(U^-_{\mathcal{S},1}(\gamma^{(\ell,m)}_c),\cdots,U^-_{\mathcal{S},m}(\gamma^{(\ell,m)}_c),W^-_i(d),i\in\Omega_{\mathcal{S}})^T$, $\mathcal{S}\in\mathcal{I}^{(\ell,m)}$. It is clear that such $V_{\mathcal{S}}$, $\mathcal{S}\in\mathcal{I}^{(\ell,m)}$, satisfy the Markov chain condition in Proposition \ref{prop:BergerTung}. Moreover,
\begin{align*}
&\mathrm{cov}^{-1}(X|V_{\mathcal{S}},\mathcal{S}\in\mathcal{I}^{(\ell,m)})\\
&=\mathrm{cov}^{-1}(X|U^-_{\mathcal{S},1}(\gamma^{(\ell,m)}_c),\cdots,U^-_{\mathcal{S},m}(\gamma^{(\ell,m)}_c),\mathcal{S}\in\mathcal{I}^{(\ell,m)})\\
&\quad+\mathrm{cov}^{-1}\left(\left(\sqrt{\frac{d^-_cd}{d^-_c-d}}Z^-_1,\cdots,\sqrt{\frac{d^-_cd}{d^-_c-d}}Z^-_{\ell}\right)^T\right)\\
&=\mathrm{diag}\left(\frac{1}{d^-_c},\cdots,\frac{1}{d^-_c}\right)+\mathrm{diag}\left(\frac{d^-_c-d}{d^-_cd},\cdots,\frac{d^-_c-d}{d^-_cd}\right)\\
&=\mathrm{diag}\left(\frac{1}{d},\cdots,\frac{1}{d}\right),
\end{align*}
which implies
\begin{align*}
\mathrm{cov}(X|V_{\mathcal{S}},\mathcal{S}\in\mathcal{I}^{(\ell,m)})=\mathrm{diag}(d,\cdots,d).
\end{align*}
Invoking Proposition \ref{prop:BergerTung} proves (\ref{eq:onedirectionnegative}) for $d\in(0,d^-_c)$.

\section{Proof of Theorem \ref{thm:theorem2}}\label{sec:proof2}

It suffices to show that
\begin{align}
r^{(\ell,m)}(d)\leq r^{(\ell,\ell)}(d),\quad d\in(0,d_c^{(\ell,m)}].\label{eq:onedirectionpositive}
\end{align}
For any $\gamma> 0$ and $\mathcal{S}\in\mathcal{I}^{(\ell,m)}$, define
\begin{align*}
U^+_{\mathcal{S}}(\gamma)\triangleq\sum\limits_{i\in\mathcal{S}}X_i+\sqrt{\gamma}N^+_{\mathcal{S}},
\end{align*}
where $N^+_{\mathcal{S}}$ is a zero-mean unit-variance Gaussian random variable. Moreover, we assume that $X$, $N^+_{\mathcal{S}}$, $\mathcal{S}\in\mathcal{I}^{(\ell,m)}$ are mutually independent. 

\begin{proposition}\label{prop:positivecoefficient}
We have
\begin{align*}
&\mathrm{cov}(X|U^+_{\mathcal{S},1}(\gamma),\cdots,U^+_{\mathcal{S},m}(\gamma),\mathcal{S}\in\mathcal{I}^{(\ell,m)})\\
&=\left(
    \begin{array}{cccc}
      d^+(\gamma) & \theta^+(\gamma) & \cdots & \theta^+(\gamma) \\
      \theta^+(\gamma) & \ddots & \ddots & \vdots \\
      \vdots & \ddots & \ddots & \theta^+(\gamma) \\
      \theta^+(\gamma) & \cdots & \theta^+(\gamma) & d^+(\gamma) \\
    \end{array}
  \right),
\end{align*}
where
\begin{align}
d^+(\gamma)&\triangleq 1-\frac{\eta_3\gamma+\eta_1}{\gamma^2+\eta_2\gamma+\eta_1},\label{eq:dplus}\\
\theta^+(\gamma)&\triangleq\rho-\frac{\eta_4\gamma+\eta_1\rho}{\gamma^2+\eta_2\gamma+\eta_1}\label{eq:thetaplus}
\end{align}
with
\begin{align*}
\eta_1&\triangleq{\ell-1\choose m-1}{\ell-2\choose m-1}m(1-\rho)(1+(\ell-1)\rho),\\
\eta_2&\triangleq{\ell-1\choose m-1}(1+(m-1)\rho)\\
&\quad+{\ell-2\choose m-1}m(1+(\ell-2)\rho)\\
&\quad+{\ell-2\choose m-2}((\ell-1)m\rho+(m-1)(1-\rho)),\\
\eta_3&\triangleq{\ell-1\choose m-1}(1+(m-1)\rho)+{\ell-2\choose m-1}(\ell-1)m\rho^2\\
&\quad+{\ell-2\choose m-2}(\ell-1)\rho(1+(m-1)\rho),\\
\eta_4&\triangleq{\ell-1\choose m-1}\rho(1+(m-1)\rho)\\
&\quad+{\ell-2\choose m-1}m\rho(1+(\ell-2)\rho)\\
&\quad+{\ell-2\choose m-2}(1+(\ell-2)\rho)(1+(m-1)\rho).
\end{align*}
\end{proposition}
\begin{IEEEproof}
See Appendix \ref{app:positivecoefficient}.
\end{IEEEproof}

Setting $\theta^+(\gamma)=0$ gives
\begin{align*}
\gamma=\gamma^{(\ell,m)}_c\triangleq\frac{{\ell-2\choose m-2}(1-\rho)(1+(\ell-1)\rho)}{\rho}.
\end{align*}
It can be verified that 
\begin{align*}
d^+(\gamma^{(\ell,m)}_c)&=1-\frac{\eta_3\gamma^{(\ell,m)}_c+\eta_1}{(\gamma^{(\ell,m)}_c)^2+\eta_2\gamma^{(\ell,m)}_c+\eta_1}\\
&=1-\frac{\eta_3\rho\gamma^{(\ell,m)}_c+\eta_1\rho}{\eta_4\gamma^{(\ell,m)}_c+\eta_1\rho}\\
&=d^{(\ell,m)}_c.
\end{align*}
Invoking Proposition \ref{prop:BergerTung} with $V_{\mathcal{S}}\triangleq U^+_{\mathcal{S}}(\gamma^{(\ell,m)}_c)$, $\mathcal{S}\in\mathcal{I}^{(\ell,m)}$, (which satisfy the Markov chain condition in Proposition \ref{prop:BergerTung}) proves (\ref{eq:onedirectionpositive}) for $d=d^{(\ell,m)}_c$.

Now consider the case $d\in(0,d^{(\ell,m)}_c)$. We will only give a sketch of the proof here since it is similar to its counterpart in Section \ref{sec:proof1}.
Let
\begin{align*}
W^+_i(d)\triangleq X_i+\sqrt{\frac{d^{(\ell,m)}_cd}{d^{(\ell,m)}_c-d}}Z^+_i,\quad i=1,\cdots,\ell,
\end{align*}
where $Z^+_1,\cdots,Z^+_{\ell}$ are mutually independent zero-mean unit variance Gaussian random variables, and are independent of $X$, $N^+_{\mathcal{S}}$, $\mathcal{S}\in\mathcal{I}^{(\ell,m)}$. Construct $\Omega_{\mathcal{S}}$, $\mathcal{S}\in\mathcal{I}^{(\ell,m)}$, such that 1) $\Omega_{\mathcal{S}}\subseteq\mathcal{S}$, $\mathcal{S}\in\mathcal{I}^{(\ell,m)}$, 2) $\Omega_{\mathcal{S}}\cap\Omega_{\mathcal{S}'}=\emptyset$, $\mathcal{S}\neq\mathcal{S}'$, 3) $\cup_{\mathcal{S}\in\mathcal{I}^{(\ell,m)}}\Omega_{\mathcal{S}}=\{1,\cdots,\ell\}$. 
Define $V_{\mathcal{S}}\triangleq(U^+_{\mathcal{S}}(\gamma^{(\ell,m)}_c),W^+_i(d),i\in\Omega_{\mathcal{S}})^T$, $\mathcal{S}\in\mathcal{I}^{(\ell,m)}$. It is clear that such $V_{\mathcal{S}}$, $\mathcal{S}\in\mathcal{I}^{(\ell,m)}$, satisfy the Markov chain condition in Proposition \ref{prop:BergerTung}, and
\begin{align*}
\mathrm{cov}(X|V_{\mathcal{S}},\mathcal{S}\in\mathcal{I}^{(\ell,m)})=\mathrm{diag}(d,\cdots,d).
\end{align*}
Invoking Proposition \ref{prop:BergerTung} proves (\ref{eq:onedirectionpositive}) for $d\in(0,d^{(\ell,m)}_c)$.

\begin{remark}\label{rmk:remark8}
Setting $d^+(\gamma)=d$ gives
\begin{align*}
\gamma&=\gamma^{(\ell,m)}\\
&\triangleq\frac{\eta_3-\eta_2(1-d)+\sqrt{(\eta_2(1-d)-\eta_3)^2+4\eta_1d(1-d)}}{2(1-d)}.
\end{align*}
Note that there is a one-to-one correspondence between $d\in(0,1)$ and $\gamma^{(\ell,m)}\in(0,\infty)$. The preceding argument in fact shows that, for $\rho\in(0,1)$ and $m=1,\cdots,\ell$,
\begin{align}
r^{(\ell,m)(d)}\leq\overline{r}^{(\ell,m)}(d),\quad d\in(0,1),\label{eq:BTupperbound}
\end{align}
where 
\begin{align*}
\overline{r}^{(\ell,m)}(d)\triangleq\frac{1}{2}\log\frac{(1-\rho)^{\ell-1}(1+(\ell-1)\rho)}{(d-\theta^{(\ell,m)})^{\ell-1}(d+(\ell-1)\theta^{(\ell,m)})}
\end{align*}
with
\begin{align}
\theta^{(\ell,m)}\triangleq\left\{
                    \begin{array}{ll}
                      0, & d\in(0,d^{(\ell,m)}_c], \\
                      \theta^+(\gamma^{(\ell,m)}), & d\in(d^{(\ell,m)}_c,1).
                    \end{array}
                  \right.\label{eq:deftheta}
\end{align}
The equality in (\ref{eq:BTupperbound}) holds for $d\in(0,d^{(\ell,m)}_c]$. Moreover, by defining ${\ell-2\choose\ell-1}\triangleq0$ and ${\ell-2\choose -1}\triangleq0$, one can readily verify that $\overline{r}^{(\ell,m)}(d)$ coincides with $r^{(\ell,m)}(d)$ for $d\in(d^{(\ell,m)}_c,1)$ when $m=\ell$ or $m=1$. However, it is still unknown whether $\overline{r}^{(\ell,m)}(d)=r^{(\ell,m)}(d)$ for $d\in(d^{(\ell,m)}_c,1)$ when $1<m<\ell$.



\end{remark}

\section{Proof of Theorem \ref{thm:theorem3}}\label{sec:proof3}

In view of Remark \ref{rmk:remark8}, Remark \ref{rmk:remark3}, and (\ref{eq:expansionell}), it suffices to show that, for $\rho\in(0,1)$ and $m\geq 1$,
\begin{align*}
&\overline{r}^{(\ell,m)}(d)=\left\{
                       \begin{array}{ll}
                         r^{(\ell,m)}_1(d), & d\in(0,d^{(m)}_c], \\
                        r^{(\ell,m)}_2(d), & d\in(d^{(m)}_c,d^+_c), \\
                       r^{(\ell,m)}_3(d) , & d=d^+_c,\\
                        r^{(\ell,m)}_4(d),  & d\in(d^+_c,1).
                       \end{array}
                     \right.
\end{align*}

First consider the case $d\in(0,d^{(m)}_c)$. When $\ell$ is sufficiently large, we have $d\in(0,d^{(\ell,m)}_c]$ and consequently 
\begin{align*}
\overline{r}^{(\ell,m)}(d)&=\frac{1}{2}\log\frac{(1-\rho)^{\ell-1}(1+(\ell-1)\rho)}{d^{\ell}}\\
&=\frac{\ell}{2}\log\frac{1-\rho}{d}+\frac{1}{2}\log\ell+\frac{1}{2}\log\frac{\rho}{1-\rho}\\
&\quad+\frac{1}{2}\log\left(1+\frac{1-\rho}{\ell\rho}\right)\\
&=r^{(\ell,m)}_1(d).
\end{align*}

Next we shall derive a few results that are needed for studying the remaining cases. It can be verified that
\begin{align*}
&\eta_1=g_1\frac{\ell^{2m}}{((m-1)!)^2}+h_1\frac{\ell^{2m-1}}{((m-1)!)^2}+O(\ell^{2m-2}),\\
&\eta_i=g_i\frac{\ell^m}{(m-1)!}+h_i\frac{\ell^{m-1}}{(m-1)!}+O(\ell^{m-2}),\quad i=2,3,4,
\end{align*}
where
\begin{align*}
&g_1\triangleq0,\quad g_2\triangleq m\rho,\quad g_3\triangleq m\rho^2,\quad g_4\triangleq m\rho^2,\\
&h_1\triangleq m\rho(1-\rho),\\
&h_2\triangleq (m+1)(1-\rho)+\frac{(m+4)m(m-1)\rho}{2},\\
&h_3\triangleq h_2\rho+(1-\rho)(1+(m-2)\rho),\\
&h_4\triangleq h_2\rho+(m-1)\rho(1-\rho).
\end{align*}
According to (\ref{eq:dplus}) and (\ref{eq:thetaplus}), 
\begin{align*}
&d=\frac{(\gamma^{(\ell,m)})^2+(\eta_2-\eta_3)\gamma^{(\ell,m)}}{(\gamma^{(\ell,m)})^2+\eta_2\gamma^{(\ell,m)}+\eta_1},\\
&\theta^+(\gamma^{(\ell,m)})=\frac{\rho(\gamma^{(\ell,m)})^2+(\eta_2\rho-\eta_4)\gamma^{(\ell,m)}}{(\gamma^{(\ell,m)})^2+\eta_2\gamma^{(\ell,m)}+\eta_1},
\end{align*}
which implies
\begin{align}
\theta^+(\gamma^{(\ell,m)})=\frac{(\rho\gamma^{(\ell,m)}+\eta_2\rho-\eta_4)d}{\gamma^{(\ell,m)}+\eta_2-\eta_3}.\label{eq:simplifiedtheta1}
\end{align}
Using the asymptotic expressions of $\eta_2$, $\eta_3$, and $\eta_4$, we can rewrite (\ref{eq:simplifiedtheta1}) as 
\begin{align}
\theta^+(\gamma^{(\ell,m)})=\frac{\rho d\gamma^{(\ell,m)}\frac{(m-1)!}{\ell^m}-\frac{(m-1)\rho(1-\rho)d}{\ell}+O(\frac{1}{\ell^2})}{\gamma^{(\ell,m)}\frac{(m-1)!}{\ell^m}+m\rho(1-\rho)+\frac{h_2-h_3}{\ell}+O(\frac{1}{\ell^2})}.\label{eq:simplifiedtheta}
\end{align}
Note that
\begin{align*}
&\eta_3-\eta_2(1-d)\\
&=m\rho(d-1+\rho)\frac{\ell^m}{(m-1)!}+(h_3-h_2(1-d))\frac{\ell^{m-1}}{(m-1)!}\\
&\quad+O(\ell^{m-2}),\\
&(\eta_2(1-d)-\eta_3)^2+4\eta_1d(1-d)\\
&=m^2\rho^2(1-\rho-d)^2\frac{\ell^{2m}}{((m-1)!)^2}+\zeta\frac{\ell^{2m-1}}{((m-1)!)^2}\\
&\quad+O(\ell^{2m-2}),
\end{align*}
where
\begin{align*}
\zeta&\triangleq 2m\rho(1-\rho-d)(h_2(1-d)-h_3)\\
&\quad+4m\rho(1-\rho)d(1-d).
\end{align*}
As a consequence,
\begin{align}
&\gamma^{(\ell,m)}\nonumber\\
&=\frac{m\rho(d-1+\rho)}{2(1-d)}\frac{\ell^m}{(m-1)!}+\frac{h_3-h_2(1-d)}{2(1-d)}\frac{\ell^{m-1}}{(m-1)!}\nonumber\\
&\quad+\frac{\sqrt{m^2\rho^2(1-\rho-d)^2+\frac{\zeta}{\ell}+O(\frac{1}{\ell^2})}}{2(1-d)}\frac{\ell^{m}}{(m-1)!}\nonumber\\
&\quad+O(\ell^{m-2}).\label{eq:restategamma}
\end{align}

Now we are in a position to study the remaining cases. 

For $d\in(0,d^+_c)$ (if $m=1$) or $d\in[d^{(m)}_c,d^+_c)$ (if $m>1$), we have
$1-\rho-d>0$. It follows from (\ref{eq:restategamma}) that
\begin{align}
&\gamma^{(\ell,m)}\nonumber\\
&=\frac{m\rho(d-1+\rho)}{2(1-d)}\frac{\ell^m}{(m-1)!}+\frac{h_3-h_2(1-d)}{2(1-d)}\frac{\ell^{m-1}}{(m-1)!}\nonumber\\
&\quad+\frac{m\rho(1-\rho-d)\sqrt{1+\frac{\zeta}{\ell m^2\rho^2(1-\rho-d)^2}+O(\frac{1}{\ell^2})}}{2(1-d)}\frac{\ell^{m}}{(m-1)!}\nonumber\\
&\quad+O(\ell^{m-2})\nonumber\\
&=\frac{m\rho(d-1+\rho)}{2(1-d)}\frac{\ell^m}{(m-1)!}+\frac{h_3-h_2(1-d)}{2(1-d)}\frac{\ell^{m-1}}{(m-1)!}\nonumber\\
&\quad+\frac{m\rho(1-\rho-d)\left(1+\frac{\zeta}{2\ell m^2\rho^2(1-\rho-d)^2}\right)}{2(1-d)}\frac{\ell^{m}}{(m-1)!}\nonumber\\
&\quad+O(\ell^{m-2})\nonumber\\
&=\frac{(1-\rho)d}{1-\rho-d}\frac{\ell^{m-1}}{(m-1)!}+O(\ell^{m-2}),\nonumber
\end{align}
which, together with (\ref{eq:simplifiedtheta}) and some simple calculation, gives
\begin{align*}
&\theta^+(\gamma^{(\ell,m)})\\
&=\frac{\frac{\rho(1-\rho)d^2}{\ell(1-\rho-d)}-\frac{(m-1)\rho(1-\rho)d}{\ell}+O(\frac{1}{\ell^2})}{m\rho(1-\rho)+O(\frac{1}{\ell})}\\
&=\left(\frac{d(d-(m-1)(1-\rho-d))}{\ell m(1-\rho-d)}+O(\frac{1}{\ell^2})\right)\left(1+O(\frac{1}{\ell})\right)\\
&=\frac{d(d-(m-1)(1-\rho-d))}{\ell m(1-\rho-d)}+O(\frac{1}{\ell^2}).
\end{align*}
One can readily verify that 
\begin{align*}
&\overline{r}^{(\ell,m)}(d)\\
&=\frac{\ell}{2}\log\frac{1-\rho}{d}+\frac{1}{2}\log\ell-\frac{\ell-1}{2}\log\left(1-\frac{\theta^+(\gamma^{(\ell,m)})}{d}\right)\\
&\quad +\frac{1}{2}\log\left(\frac{\rho d}{1-\rho}+\frac{d}{\ell}\right)-\frac{1}{2}\log(d+(\ell-1)\theta^+(\gamma^{(\ell,m)}))\\
&=r^{(\ell,m)}_2(d).
\end{align*}

For $d=d^+_c$, we have $1-\rho-d=0$. It follows from (\ref{eq:restategamma}) that
\begin{align}
&\gamma^{(\ell,m)}\nonumber\\
&=\frac{h_3-h_2(1-d)}{2(1-d)}\frac{\ell^{m-1}}{(m-1)!}\nonumber\\
&\quad+\frac{\sqrt{\frac{4m\rho(1-\rho)d(1-d)}{\ell}+O(\frac{1}{\ell^2})}}{2(1-d)}\frac{\ell^m}{(m-1)!}+O(\ell^{m-2})\nonumber\\
&=\sqrt{m}(1-\rho)\frac{\ell^{m-\frac{1}{2}}}{(m-1)!}+\frac{h_3-h_2\rho}{2\rho}\frac{\ell^{m-1}}{(m-1)!}+O(\ell^{m-\frac{3}{2}})\nonumber\\
&=\sqrt{m}(1-\rho)\frac{\ell^{m-\frac{1}{2}}}{(m-1)!}+\frac{(1-\rho)(1+(m-2)\rho)}{2\rho}\frac{\ell^{m-1}}{(m-1)!}\nonumber\\
&\quad+O(\ell^{m-\frac{3}{2}}),\nonumber
\end{align}
which, together with (\ref{eq:simplifiedtheta}) and some simple calculation, gives
\begin{align*}
&\theta^+(\gamma^{(\ell,m)})\\
&=\frac{\frac{\sqrt{m}\rho(1-\rho)^2}{\sqrt{\ell}}+\frac{(1-\rho)^2(1+(m-2)\rho-2(m-1)\rho)}{2\ell}+O(\frac{1}{\ell^2})}
{m\rho(1-\rho)+\frac{\sqrt{m}(1-\rho)}{\sqrt{\ell}}+O(\frac{1}{\ell})}\\
&=\left(\frac{1-\rho}{\sqrt{\ell m}}+\frac{(1-\rho)(1-m\rho)}{2\ell m\rho}+O(\frac{1}{\ell^2})\right)\\
&\quad\times\left(1-\frac{1}{\sqrt{\ell m}\rho}+O(\frac{1}{\ell})\right)\\
&=\frac{1-\rho}{\sqrt{\ell m}}-\frac{(1-\rho)(1+m\rho)}{2\ell m\rho}+O(\frac{1}{\ell^{\frac{3}{2}}}).
\end{align*}
One can readily verify that 
\begin{align*}
\overline{r}^{(\ell,m)}(d)&=-\frac{\ell-1}{2}\log\left(1-\frac{\theta^+(\gamma^{(\ell,m)})}{1-\rho}\right)+\frac{1}{4}\log\ell\\
&\quad+\frac{1}{2}\log\left(\rho+\frac{1-\rho}{\ell}\right)\\
&\quad-\frac{1}{2}\log\left(\frac{1-\rho+(\ell-1)\theta^+(\gamma^{(\ell,m)})}{\sqrt{\ell}}\right)\\
&=r^{(\ell,m)}_3(d).
\end{align*}

For $d\in(d^+_c,1)$, we have $1-\rho-d<0$. It follows from (\ref{eq:restategamma}) that
\begin{align}
&\gamma^{(\ell,m)}\nonumber\\
&=\frac{m\rho(d-1+\rho)}{2(1-d)}\frac{\ell^m}{(m-1)!}+\frac{h_3-h_2(1-d)}{2(1-d)}\frac{\ell^{m-1}}{(m-1)!}\nonumber\\
&\quad+\frac{m\rho(d-1+\rho)\sqrt{1+\frac{\zeta}{\ell m^2\rho^2(1-\rho-d)^2}+O(\frac{1}{\ell^2})}}{2(1-d)}\frac{\ell^{m}}{(m-1)!}\nonumber\\
&\quad+O(\ell^{m-2})\nonumber\\
&=\frac{m\rho(d-1+\rho)}{2(1-d)}\frac{\ell^m}{(m-1)!}+\frac{h_3-h_2(1-d)}{2(1-d)}\frac{\ell^{m-1}}{(m-1)!}\nonumber\\
&\quad+\frac{m\rho(d-1+\rho)\left(1+\frac{\zeta}{2\ell m^2\rho^2(1-\rho-d)^2}\right)}{2(1-d)}\frac{\ell^{m}}{(m-1)!}\nonumber\\
&\quad+O(\ell^{m-2})\nonumber\\
&=\frac{m\rho(d-1+\rho)}{1-d}\frac{\ell^m}{(m-1)!}\nonumber\\
&\quad+\left(\frac{h_3-h_2(1-d)}{1-d}+\frac{(1-\rho)d}{d-1+\rho}\right)\frac{\ell^{m-1}}{(m-1)!}+O(\ell^{m-2}).\label{eq:case4sub}
\end{align} 
Substituting (\ref{eq:case4sub}) into (\ref{eq:simplifiedtheta}) gives
\begin{align*}
\theta^+(\gamma^{(\ell,m)})=\frac{d-1+\rho+\frac{\mu}{\ell}+O(\frac{1}{\ell^2})}{1+\frac{\nu}{\ell}+O(\frac{1}{\ell^2})},
\end{align*}
where
\begin{align*}
\mu&\triangleq\frac{h_3-h_2(1-d)}{m\rho}+\frac{(1-\rho)d(1-d)}{m\rho(d-1+\rho)}\\
&\quad-\frac{(m-1)(1-\rho)(1-d)}{m\rho},\\
\nu&\triangleq\frac{h_3}{m\rho^2}+\frac{(1-\rho)(1-d)}{m\rho^2(d-1+\rho)}.
\end{align*}
Clearly, we have
\begin{align*}
&\theta^+(\gamma^{(\ell,m)})\\
&=\left(d-1+\rho+\frac{\mu}{\ell}+O(\frac{1}{\ell^2})\right)\left(1-\frac{\nu}{\ell}+O(\frac{1}{\ell^2})\right)\\
&=d-1+\rho+\frac{\mu-(d-1+\rho)\nu}{\ell}+O(\frac{1}{\ell^2})\\
&=d-1+\rho+\left(\frac{h_3-h_2(1-d)}{\ell m\rho}+\frac{(1-\rho)d(1-d)}{\ell m\rho(d-1+\rho)}\right.\\
&\quad-\frac{(m-1)(1-\rho)(1-d)}{\ell m\rho}-\frac{h_3(d-1+\rho)}{\ell m\rho^2}\\
&\quad\left.-\frac{(1-\rho)(1-d)}{\ell m\rho^2}\right)+O(\frac{1}{\ell^2})\\
&=d-1+\rho+\left(\frac{(h_3-h_2\rho)(1-d)}{\ell m\rho^2}+\frac{(1-\rho)d(1-d)}{\ell m\rho(d-1+\rho)}\right.\\
&\quad\left.-\frac{(m-1)(1-\rho)(1-d)}{\ell m\rho}-\frac{(1-\rho)(1-d)}{\ell m\rho^2}\right)\\
&\quad+O(\frac{1}{\ell^2})\\
&=d-1+\rho+\left(\frac{(1-\rho)(1+(m-2)\rho)(1-d)}{\ell m\rho^2}\right.\\
&\quad+\frac{(1-\rho)d(1-d)}{\ell m\rho(d-1+\rho)}-\frac{(m-1)(1-\rho)(1-d)}{\ell m\rho}\\
&\quad\left.-\frac{(1-\rho)(1-d)}{\ell m\rho^2}\right)+O(\frac{1}{\ell^2})\\
&=d-1+\rho+\frac{(1-\rho)^2(1-d)}{\ell m\rho(d-1+\rho)}+O(\frac{1}{\ell^2}).
\end{align*}
One can readily verify that
\begin{align*}
\overline{r}^{(\ell,m)}(d)&=\frac{1}{2}\log\frac{1+(\ell-1)\rho}{d+(\ell-1)\theta^+(\gamma^{(\ell,m)})}\\
&\quad-\frac{\ell-1}{2}\log\frac{d-\theta^+(\gamma^{(\ell,m)})}{1-\rho}\\
&=r^{(\ell,m)}_4(d).
\end{align*}
This completes the proof of Theorem \ref{thm:theorem3}.

\section{Numerical Results}\label{sec:numerical}

Some numerical examples will be provided in this section to illustrate our main results. We focus on the case $\rho>0$ since, in view of Theorem \ref{thm:theorem1}, the relevant plots are not particularly interesting when $\rho\leq 0$.

\begin{figure}[tb]
\hspace{-0.3in}
\includegraphics[width=10cm]{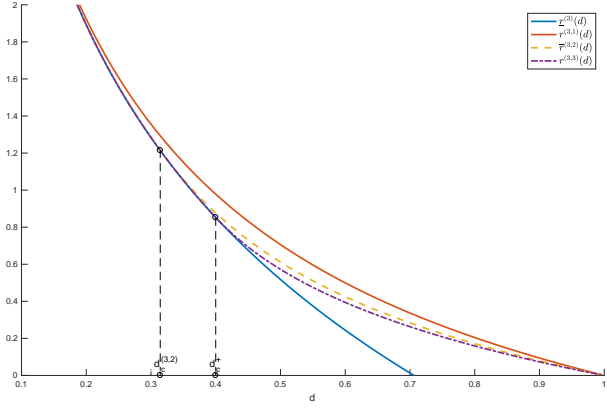}
\caption{An illustration of $\underline{r}^{(3)}(d)$, $r^{(3,1)}(d)$, $\overline{r}^{(3,2)}(d)$, and $r^{(3,3)}(d)$ with $\rho=0.6$. \label{fig:rd1}}
\end{figure}

\begin{figure}[tb]
\hspace{-0.3in}
\includegraphics[width=10cm]{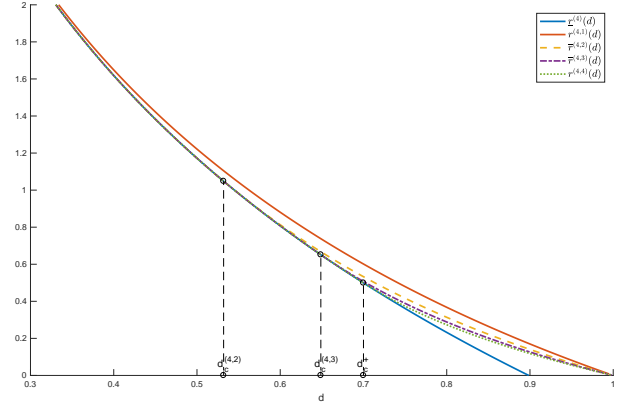}
\caption{An illustration of $\underline{r}^{(4)}(d)$, $r^{(4,1)}(d)$, $\overline{r}^{(4,2)}(d)$, $\overline{r}^{(4,3)}(d)$, and $r^{(4,4)}(d)$ with $\rho=0.3$.  \label{fig:rd2}}
\end{figure}

First we compare $\overline{r}^{(\ell,m)}(d)$ (the best known upper bound on $r^{(\ell,m)}(d)$), $1<m<\ell$, with $r^{(\ell,\ell)}(d)$ (the rate-distortion function in the centralized setting), $r^{(\ell,1)}(d)$ (the rate-distortion function in the distributed setting), and $\underline{r}^{(\ell)}(d)$ (the Shannon lower bound).
Fig. \ref{fig:rd1} illustrates the case $\ell=3$ with $\rho=0.6$. It can be seen that $r^{(3,3)}(d)$ coincides with $\underline{r}^{(3)}(d)$ when $d\leq d^+_c=0.4$, and $\overline{r}^{(3,2)}(d)$ coincides with $r^{(3,3)}(d)$ as well as $\underline{r}^{(3)}(d)$ when $d\leq d^{(3,2)}_c=\frac{11}{35}\approx0.314$. On the other hand, $r^{(3,1)}(d)$ is strictly above all the other curves for $d\in(0,1)$. See a similar plot for the case $\ell=4$ with $\rho=0.3$ in Fig. \ref{fig:rd2}, where
$d^+_c=0.7$, $d^{(4,2)}=0.532$, and $d^{(4,3)}=\frac{133}{205}\approx0.649$.

\begin{figure}[tb]
\hspace{-0.3in}
\includegraphics[width=10cm]{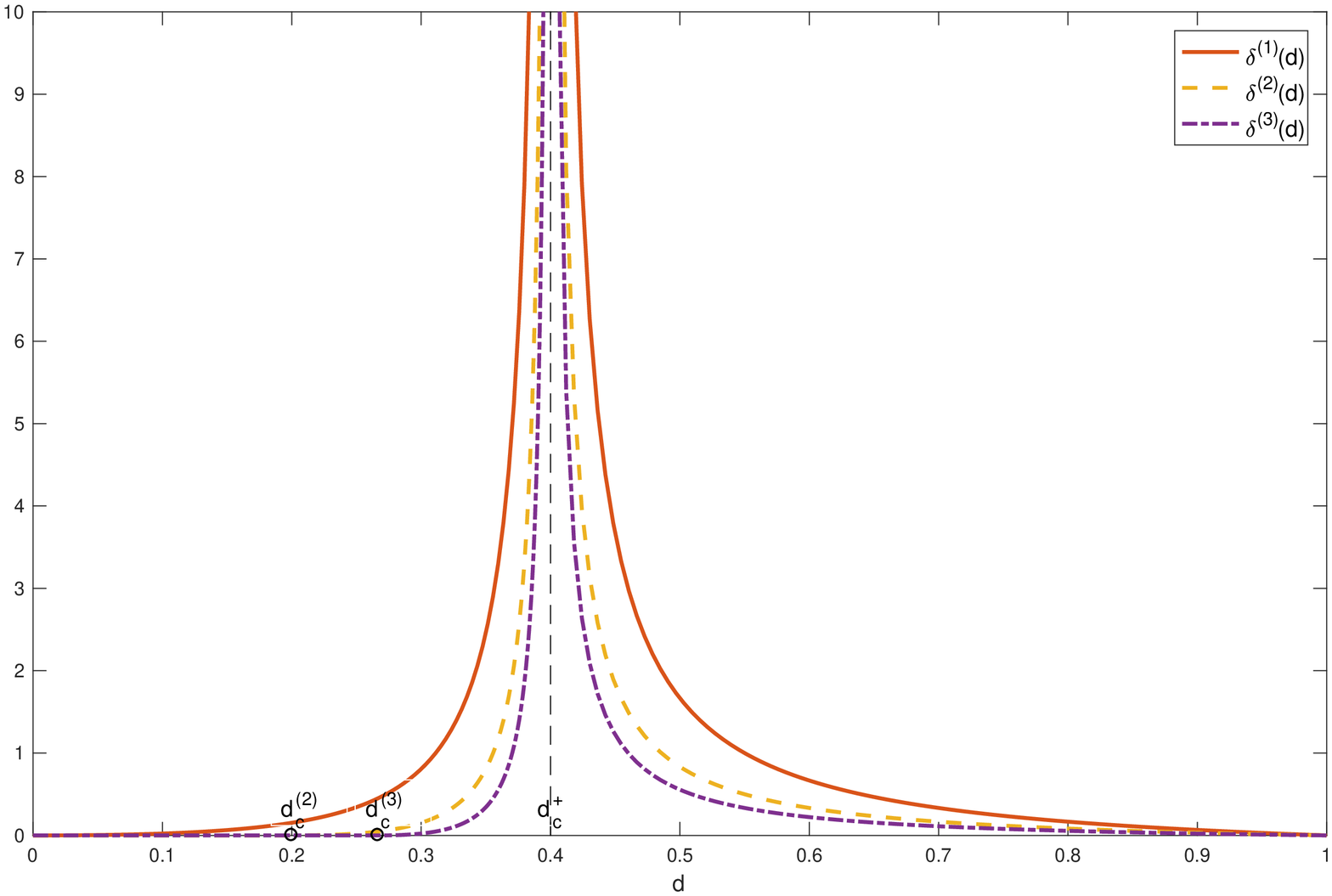}
\caption{An illustration of $\delta^{(1)}(d)$, $\delta^{(2)}(d)$, and $\delta^{(3)}(d)$ with $\rho=0.6$. \label{fig:gap1}}
\end{figure}

\begin{figure}[tb]
\hspace{-0.3in}
\includegraphics[width=10cm]{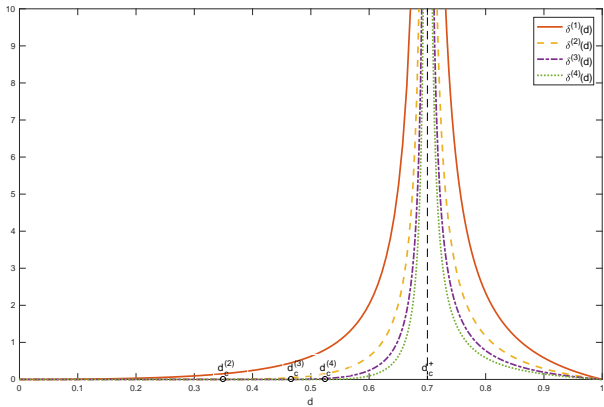}
\caption{An illustration of $\delta^{(1)}(d)$, $\delta^{(2)}(d)$, $\delta^{(3)}(d)$, and $\delta^{(4)}(d)$ with $\rho=0.3$. \label{fig:gap2}}
\end{figure}

Next we compare $\delta^{(m)}(d)$ for different values of $m$. Note that $\delta^{(m)}(d)$ indicates the asymptotic gap between $\overline{r}^{(\ell,m)}(d)$ and $r^{(\ell,\ell)}(d)$ in the large $\ell$ limit. Fig. \ref{fig:gap1} provides an illustration of $\delta^{(1)}(d)$, $\delta^{(2)}(d)$, and $\delta^{(3)}(d)$ with $\rho=0.6$. It can be seen that all the curves blow up at at the critical distortion $d^+_c=0.4$. Moreover, we have $\delta^{(2)}(d)=0$ when $d\leq d^{(2)}_c=0.2$, and $\delta^{(3)}(d)=0$ when $d\leq d^{(3)}_c=\frac{4}{15}\approx0.267$. On the other hand, $\delta^{(1)}(d)$ is strictly above zero for $d\in(0,1)$. See also a plot of $\delta^{(1)}(d)$, $\delta^{(2)}(d)$, $\delta^{(3)}(d)$, and $\delta^{(4)}(d)$ with $\rho=0.3$ in Fig. \ref{fig:gap2}, where $d^+_c=0.7$, $d^{(2)}_c=0.35$, $d^{(3)}_c=\frac{7}{15}\approx0.467$, and $d^{(4)}_c=0.525$.



\begin{figure}[tb]
\hspace{-0.3in}
\includegraphics[width=10cm]{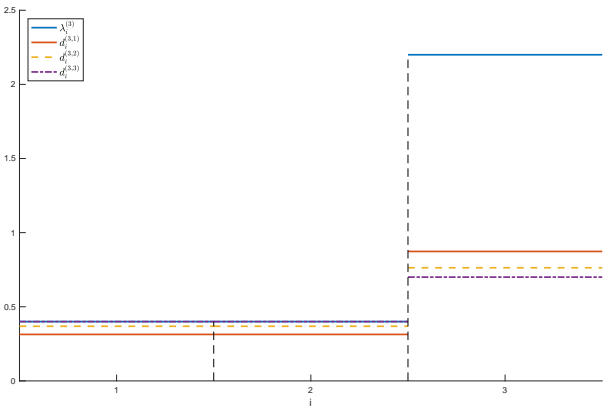}
\caption{An illustration of $\lambda^{(3)}_i$, $d^{(3,1)}_i$, $d^{(3,2)}_i$, and $d^{(3,3)}_i$, $i=1,2,3$, with $\rho=0.6$ and $d=0.5$. \label{fig:bar1}}
\end{figure}

\begin{figure}[tb]
\hspace{-0.3in}
\includegraphics[width=10cm]{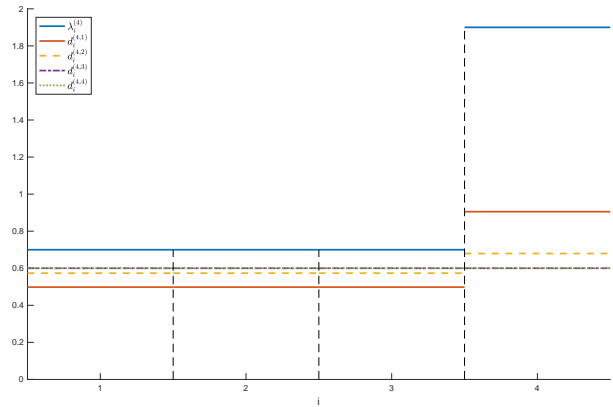}
\caption{An illustration of $\lambda^{(4)}_i$, $d^{(4,1)}_i$, $d^{(4,2)}_i$, $d^{(4,3)}_i$, and $d^{(4,4)}_i$, $i=1,2,3,4$, with $\rho=0.3$ and $d=0.6$. \label{fig:bar2}}
\end{figure}

Finally we shall perform comparisons in the eigenspace. Define
\begin{align*}
D^{(\ell,m)}\triangleq\left(
  \begin{array}{cccc}
    d & \theta^{(\ell,m)} & \cdots & \theta^{(\ell,m)} \\
    \theta^{(\ell,m)} & \ddots & \ddots &   \vdots \\
    \vdots & \ddots & \ddots & \theta^{(\ell,m)} \\
     \theta^{(\ell,m)} & \cdots & \theta^{(\ell,m)} & d \\
  \end{array}
\right),
\end{align*}
where $\theta^{(\ell,m)}$ is given by (\ref{eq:deftheta}). One can interpret as $D^{(\ell,m)}$ the  distortion covariance matrix associated with $\overline{r}^{(\ell,m)}(d)$. Indeed, we have
\begin{align*}
\overline{r}^{(\ell,m)}(d)=\frac{1}{2}\log\frac{\det(\Sigma^{(\ell)})}{\det(D^{(\ell,m)})}
\end{align*}
or equivalently
\begin{align*}
\overline{r}^{(\ell,m)}(d)=\sum\limits_{i=1}^{\ell}\frac{1}{2}\log\frac{\lambda^{(\ell)}_i}{d^{(\ell,m)}_i},
\end{align*}
where
\begin{align*}
&d^{(\ell,m)}_i\triangleq d-\theta^{(\ell,m)},\quad i=1,\cdots,\ell-1,\\
&d^{(\ell,m)}_{\ell}\triangleq d+(\ell-1)\theta^{(\ell,m)}
\end{align*}
are the eigenvalues of $D^{(\ell,m)}$. Note that $(d^{(\ell,\ell)}_1,\cdots,d^{(\ell,\ell)}_{\ell})$ corresponds to the reverse water-filling solution. Fig. \ref{fig:bar1} provides an illustration of $\lambda^{(3)}_i$, $d^{(3,1)}_i$, $d^{(3,2)}_i$, and $d^{(3,3)}_i$, $i=1,2,3$, with $\rho=0.6$ and $d=0.5$. Since $d^+_c=0.4<d$, the reverse water-filling solution leaves some dimensions uncoded; indeed, it can be seen that $d^{(3,3)}_i=\lambda^{(3)}_i$, $i=1,2$. In contrast,  for $m=1$ and $m=2$, we have $d^{(3,m)}_i<\lambda^{(3)}_i$, $i=1,2,3$, and consequently all dimensions are coded, which is suboptimal as compared to the reverse water-filling solution; nevertheless, increasing from $m=1$ to $m=2$ gets $(d^{(3,m)}_1,d^{(3,m)}_2,d^{(3,m)}_3)$ closer to the reverse water-filling solution, resulting in an improved rate-distortion performance. Fig. \ref{fig:bar2} depicts $\lambda^{(4)}_i$, $d^{(4,1)}_i$, $d^{(4,2)}_i$, $d^{(4,3)}_i$, and $d^{(4,4)}_i$, $i=1,2,3,4$, with $\rho=0.3$ and $d=0.6$. Since $d^{(4,3)}_c\approx0.649>d$, it follows that $(d^{(4,3)}_1,d^{(4,3)}_2,d^{(4,3)}_3,d^{(4,3)}_4)$ coincides with $(d^{(4,4)}_1,d^{(4,4)}_2,d^{(4,4)}_3,d^{(4,4)}_4)$. That is to say, for such $d$, the encoders in a $(4,3)$ generalized multiterminal source coding system can achieve the same effect as that of the reverse water-filling solution in the centralized setting even though they cannot fully cooperate.






\section{Conclusion}\label{sec:conclusion}

We have studied the rate-distortion limit of generalized multiterminal source coding of symmetrically correlated Gaussian sources. Although a complete characterization of this limit has been obtained when the correlation coefficient is non-positive, a lot remains to be done for the positive correlation coefficient case. We conjecture that the upper bound established in the present work, i.e., $\overline{r}^{(\ell,m)}(d)$, is tight even when $d$ is greater than $d^{(\ell,m)}_c$. However, a rigorous proof of this conjecture (even in the large $\ell$ limit) is likely to be non-trivial and may require new techniques yet to be developed.

We would like to mention that the proof of Theorems \ref{thm:theorem1} and \ref{thm:theorem2} was partly inspired by the consideration of the graphical model (more precisely, the Markov network) of a symmetric multivariate Gaussian distribution. It is of considerable interest to know whether a more conceptual proof can be constructed along that line. Moreover, probabilistic graphical models are expected to play an essential role in identifying the non-Gaussian counterpart of our problem and establishing the corresponding results.






%

\appendices
\section{Proof of Proposition \ref{prop:negativecoefficient}}\label{app:negativecoefficient}

Let $\hat{X}^-_i(\gamma)\triangleq\mathbb{E}[X_i|U^-_{\mathcal{S},1}(\gamma),\cdots,U^-_{\mathcal{S},m}(\gamma),\mathcal{S}\in\mathcal{I}^{(\ell,m)}]$, $i=1,\cdots,\ell$. We shall first prove that
\begin{align*}
\hat{X}^-_i(\gamma)=\kappa\sum\limits_{\mathcal{S}\in\mathcal{I}^{(\ell,m)}:i\in\mathcal{S}}U^-_{\mathcal{S},\tau(i)}(\gamma),\quad i=1,\cdots,\ell,
\end{align*}
where $\tau(i)$ indicates the position of $i$ in $\mathcal{S}$ when the elements of $\mathcal{S}$ are arranged in ascending order, and
\begin{align*}
\kappa\triangleq\frac{(1-\rho)}{\gamma+{\ell-2\choose m-2}\ell(1-\rho)}.
\end{align*}
It suffices to verify that, for any $\mathcal{S}'\in\mathcal{I}^{(\ell,m)}$ and $i'\in\mathcal{S}'$,
\begin{align}
&\mathbb{E}\left[\left(X_i-\kappa\sum\limits_{\mathcal{S}\in\mathcal{I}^{(\ell,m)}:i\in\mathcal{S}}U^-_{\mathcal{S},\tau(i)}(\gamma)\right)U^-_{\mathcal{S}',\tau(i')}(\gamma)\right]=0,\nonumber\\
&\hspace{2in} i=1,\cdots,\ell.\label{eq:orth}
\end{align}
Note that
\begin{align}
&X_i-\kappa\sum\limits_{\mathcal{S}\in\mathcal{I}^{(\ell,m)}:i\in\mathcal{S}}U^-_{\mathcal{S},\tau(i)}(\gamma)\nonumber\\
&=\left(1-\kappa{\ell-2\choose m-2}\ell\right)X_i+\kappa{\ell-2\choose m-2}\sum\limits_{j=1}^{\ell}X_j\nonumber\\
&\quad-\kappa\sqrt{\gamma}\sum\limits_{\mathcal{S}\in\mathcal{I}^{(\ell,m)}:i\in\mathcal{S}}N^-_{\mathcal{S},\tau(i)}.\label{eq:alternative}
\end{align}
One can readily compute that
\begin{align}
&\mathbb{E}[X_iU^-_{\mathcal{S}',\tau(i')}(\gamma)]=\left\{
                                                     \begin{array}{ll}
                                                      (m-1)(1-\rho), & i=i', \\
                                                       -(1-\rho), & i\in\mathcal{S}', i\neq i',\\
                                                       0, & i\notin\mathcal{S}',
                                                     \end{array}
                                                   \right.\label{eq:sub1}\\
&\sum\limits_{j=1}^{\ell}\mathbb{E}[X_jU^-_{\mathcal{S}',\tau(i')}(\gamma)]=0,\label{eq:sub2}\\
&\sum\limits_{\mathcal{S}\in\mathcal{I}^{(\ell,m)}:i\in\mathcal{S}}\mathbb{E}[N^-_{\mathcal{S},\tau(i)}U^-_{\mathcal{S}',\tau(i')}(\gamma)]\nonumber\\
&=\left\{
    \begin{array}{ll}
      (m-1)\sqrt{\gamma}, & i=i', \\
      -\sqrt{\gamma}, & i\in\mathcal{S}', i\neq i', \\
      0, & i\notin\mathcal{S}'.
    \end{array}
  \right.\label{eq:sub3}
\end{align}
Combining (\ref{eq:alternative}), (\ref{eq:sub1}), (\ref{eq:sub2}), and (\ref{eq:sub3}) gives (\ref{eq:orth}).

For $i=1,\cdots,\ell$,
\begin{align}
&\mathbb{E}[(X_i-\hat{X}^-_i(\gamma))^2]\nonumber\\
&=\mathbb{E}[(X_i-\hat{X}^-_i(\gamma))X_i]-\mathbb{E}[(X_i-\hat{X}^-_i(\gamma))\hat{X}^-_i(\gamma)]\nonumber\\
&=\mathbb{E}[(X_i-\hat{X}^-_i(\gamma))X_i]\label{eq:dueto1}\\
&=\left(1-\kappa{\ell-2\choose m-2}\ell\right)\mathbb{E}[X^2_i]\nonumber\\
&\quad+\kappa{\ell-2\choose m-2}\sum\limits_{j=1}^{\ell}\mathbb{E}[X_jX_i]\nonumber\\
&\quad-\kappa\sqrt{\gamma}\sum\limits_{\mathcal{S}\in\mathcal{I}^{(\ell,m)}:i\in\mathcal{S}}\mathbb{E}[N^-_{\mathcal{S},\tau(i)}X_i]\label{eq:dueto2}\\
&=1-\kappa{\ell-2\choose m-2}\ell+\kappa{\ell-2\choose m-2}(1+(\ell-1)\rho)\nonumber\\
&=d^-(\gamma),\nonumber
\end{align}
where (\ref{eq:dueto1}) and (\ref{eq:dueto2}) are due to (\ref{eq:orth}) and (\ref{eq:alternative}), respectively. 
Moreover, for $i,i'\in\{1,\cdots,\ell\}$ with $i\neq i'$,
\begin{align}
&\mathbb{E}[(X_i-\hat{X}^-_i(\gamma))(X_{i'}-\hat{X}^-_{i'}(\gamma))]\nonumber\\
&=\mathbb{E}[(X_i-\hat{X}^-_i(\gamma))X_{i'}]-\mathbb{E}[(X_i-\hat{X}^-_i(\gamma))\hat{X}^-_{i'}(\gamma)]\nonumber\\
&=\mathbb{E}[(X_i-\hat{X}^-_i(\gamma))X_{i'}]\label{eq:againdueto1}\\
&=\left(1-\kappa{\ell-2\choose m-2}\ell\right)\mathbb{E}[X_iX'_i]\nonumber\\
&\quad+\kappa{\ell-2\choose m-2}\sum\limits_{j=1}^{\ell}\mathbb{E}[X_jX_{i'}]\nonumber\\
&\quad-\kappa\sqrt{\gamma}\sum\limits_{\mathcal{S}\in\mathcal{I}^{(\ell,m)}:i\in\mathcal{S}}\mathbb{E}[N^-_{\mathcal{S},\tau(i)}X_{i'}]\label{eq:againdueto2}\\
&=\rho-\kappa{\ell-2\choose m-2}\ell\rho+\kappa{\ell-2\choose m-2}(1+(\ell-1)\rho)\nonumber\\
&=\theta^-(\gamma),\nonumber
\end{align}
where (\ref{eq:againdueto1}) and (\ref{eq:againdueto2}) are due to (\ref{eq:orth}) and (\ref{eq:alternative}), respectively. This completes the proof of Proposition \ref{prop:negativecoefficient}.

\section{Proof of Proposition \ref{prop:positivecoefficient}}\label{app:positivecoefficient}

Let $\hat{X}^+_i(\gamma)\triangleq\mathbb{E}[X_i|U^+_{\mathcal{S}}(\gamma),\mathcal{S}\in\mathcal{I}^{(\ell,m)}]$, $i=1,\cdots,\ell$. We shall first prove that
\begin{align*}
&\hat{X}^+_i(\gamma)=\alpha\sum\limits_{\mathcal{S}\in\mathcal{I}^{(\ell,m)}:i\in\mathcal{S}}U^+_{\mathcal{S}}(\gamma)+\beta\sum\limits_{\mathcal{S}\in\mathcal{I}^{(\ell,m)}:i\notin\mathcal{S}}U^+_{\mathcal{S}}(\gamma),\\
&\hspace{2in} i=1,\cdots,\ell,
\end{align*}
where
\begin{align*}
&\alpha\triangleq\frac{(1+(m-1)\rho)\gamma+{\ell-2\choose m-1}m(1-\rho)(1+(\ell-1)\rho)}{\gamma^2+\eta_2\gamma+\eta_1},\\
&\beta\triangleq\frac{m\rho\gamma-{\ell-2\choose m-2}m(1-\rho)(1+(\ell-1)\rho)}{\gamma^2+\eta_2\gamma+\eta_1}.
\end{align*}
It suffices to verify that, for any $\mathcal{S}'\in\mathcal{I}^{(\ell,m)}$,
\begin{align}
&\mathbb{E}\left[\left(X_i-\alpha\sum\limits_{\mathcal{S}\in\mathcal{I}^{(\ell,m)}:i\in\mathcal{S}}U^+_{\mathcal{S}}(\gamma)\right.\right.\nonumber\\
&\left.\left.\qquad-\beta\sum\limits_{\mathcal{S}\in\mathcal{I}^{(\ell,m)}:i\notin\mathcal{S}}U^+_{\mathcal{S}}(\gamma)\right)U^+_{\mathcal{S}'}(\gamma)\right]=0,\quad i=1,\cdots,\ell.\label{eq:orthpositive}
\end{align}
Note that
\begin{align}
&X_i-\alpha\sum\limits_{\mathcal{S}\in\mathcal{I}^{(\ell,m)}:i\in\mathcal{S}}U^+_{\mathcal{S}}(\gamma)-\beta\sum\limits_{\mathcal{S}\in\mathcal{I}^{(\ell,m)}:i\notin\mathcal{S}}U^+_{\mathcal{S}}(\gamma)\nonumber\\
&=\left(1-\alpha{\ell-1\choose m-1}+\alpha{\ell-2\choose m-2}+\beta{\ell-2\choose m-1}\right)X_i\nonumber\\
&\quad-\left(\alpha{\ell-2\choose m-2}+\beta{\ell-2\choose m-1}\right)\sum\limits_{j=1}^{\ell}X_j\nonumber\\
&\quad-(\alpha-\beta)\sqrt{\gamma}\sum\limits_{\mathcal{S}\in\mathcal{I}^{(\ell,m)}:i\in\mathcal{S}}N^+_{\mathcal{S}}\nonumber\\
&\quad-\beta\sqrt{\gamma}\sum\limits_{\mathcal{S}\in\mathcal{I}^{(\ell,m)}}N^+_{\mathcal{S}}.\label{eq:alternativepositive}
\end{align}
One can readily compute that
\begin{align}
&\mathbb{E}[X_iU^+_{\mathcal{S}'}(\gamma)]=\left\{
                                                     \begin{array}{ll}
                                                       1+(m-1)\rho, & i\in\mathcal{S}', \\
                                                       m\rho, & i\notin\mathcal{S}',
                                                     \end{array}
                                                   \right.\label{eq:sub1positive}\\
&\sum\limits_{j=1}^{\ell}\mathbb{E}[X_jU^+_{\mathcal{S}'}(\gamma)]=m(1+(\ell-1)\rho),\label{eq:sub2positive}\\
&\sum\limits_{\mathcal{S}\in\mathcal{I}^{(\ell,m)}:i\in\mathcal{S}}\mathbb{E}[N^+_{\mathcal{S}}U^+_{\mathcal{S}'}(\gamma)]=\left\{
    \begin{array}{ll}
      \sqrt{\gamma}, & i\in\mathcal{S}', \\
      0, & i\notin\mathcal{S}',
    \end{array}
  \right.\label{eq:sub3positive}\\
&\sum\limits_{\mathcal{S}\in\mathcal{I}^{(\ell,m)}}\mathbb{E}[N^+_{\mathcal{S}}U^+_{\mathcal{S}'}(\gamma)]=\sqrt{\gamma}.\label{eq:sub4positive}
\end{align}
Combining (\ref{eq:alternativepositive}), (\ref{eq:sub1positive}), (\ref{eq:sub2positive}), (\ref{eq:sub3positive}), and (\ref{eq:sub4positive}) gives (\ref{eq:orthpositive}).

For $i=1,\cdots,\ell$,
\begin{align}
&\mathbb{E}[(X_i-\hat{X}^+_i(\gamma))^2]\nonumber\\
&=\mathbb{E}[(X_i-\hat{X}^+_i(\gamma))X_i]-\mathbb{E}[(X_i-\hat{X}^+_i(\gamma))\hat{X}^+_i(\gamma)]\nonumber\\
&=\mathbb{E}[(X_i-\hat{X}^+_i(\gamma))X_i]\label{eq:dueto1positive}\\
&=\left(1-\alpha{\ell-1\choose m-1}+\alpha{\ell-2\choose m-2}+\beta{\ell-2\choose m-1}\right)\mathbb{E}[X^2_i]\nonumber\\
&\quad-\left(\alpha{\ell-2\choose m-2}+\beta{\ell-2\choose m-1}\right)\sum\limits_{j=1}^{\ell}\mathbb{E}[X_jX_i]\nonumber\\
&\quad-(\alpha-\beta)\sqrt{\gamma}\sum\limits_{\mathcal{S}\in\mathcal{I}^{(\ell,m)}:i\in\mathcal{S}}\mathbb{E}[N^+_{\mathcal{S}}X_i]\nonumber\\
&\quad-\beta\sqrt{\gamma}\sum\limits_{\mathcal{S}\in\mathcal{I}^{(\ell,m)}}\mathbb{E}[N^+_{\mathcal{S}}X_i]\label{eq:dueto2positive}\\
&=1-\alpha{\ell-1\choose m-1}+\alpha{\ell-2\choose m-2}+\beta{\ell-2\choose m-1}\nonumber\\
&\quad-\left(\alpha{\ell-2\choose m-2}+\beta{\ell-2\choose m-1}\right)(1+(\ell-1)\rho)\nonumber\\
&=d^+(\gamma),\nonumber
\end{align}
where (\ref{eq:dueto1positive}) and (\ref{eq:dueto2positive}) are due to (\ref{eq:orthpositive}) and (\ref{eq:alternativepositive}), respectively. 
Moreover, for $i,i'\in\{1,\cdots,\ell\}$ with $i\neq i'$,
\begin{align}
&\mathbb{E}[(X_i-\hat{X}^+_i(\gamma))(X_{i'}-\hat{X}^+_{i'}(\gamma))]\nonumber\\
&=\mathbb{E}[(X_i-\hat{X}^+_i(\gamma))X_{i'}]-\mathbb{E}[(X_i-\hat{X}^+_i(\gamma))\hat{X}^+_{i'}(\gamma)]\nonumber\\
&=\mathbb{E}[(X_i-\hat{X}^+_i(\gamma))X_{i'}]\label{eq:againdueto1positive}\\
&=\left(1-\alpha{\ell-1\choose m-1}+\alpha{\ell-2\choose m-2}+\beta{\ell-2\choose m-1}\right)\mathbb{E}[X_iX_{i'}]\nonumber\\
&\quad-\left(\alpha{\ell-2\choose m-2}+\beta{\ell-2\choose m-1}\right)\sum\limits_{j=1}^{\ell}\mathbb{E}[X_jX_{i'}]\nonumber\\
&\quad-(\alpha-\beta)\sqrt{\gamma}\sum\limits_{\mathcal{S}\in\mathcal{I}^{(\ell,m)}:i\in\mathcal{S}}\mathbb{E}[N^+_{\mathcal{S}}X_{i'}]\nonumber\\
&\quad-\beta\sqrt{\gamma}\sum\limits_{\mathcal{S}\in\mathcal{I}^{(\ell,m)}}\mathbb{E}[N^+_{\mathcal{S}}X_{i'}]\label{eq:againdueto2positive}\\
&=\rho-\alpha{\ell-1\choose m-1}\rho+\alpha{\ell-2\choose m-2}\rho+\beta{\ell-2\choose m-1}\rho\nonumber\\
&\quad-\left(\alpha{\ell-2\choose m-2}+\beta{\ell-2\choose m-1}\right)(1+(\ell-1)\rho)\nonumber\\
&=\theta^+(\gamma),\nonumber
\end{align}
where (\ref{eq:againdueto1positive}) and (\ref{eq:againdueto2positive}) are due to (\ref{eq:orthpositive}) and (\ref{eq:alternativepositive}), respectively. This completes the proof of Proposition \ref{prop:positivecoefficient}.



\ifCLASSOPTIONcaptionsoff
  \newpage
\fi

\end{document}